\documentclass{sig-alternate}
\usepackage[boxed,linesnumbered,figure]{algorithm2e}
\usepackage{tweaklist}
\usepackage{array}
\usepackage{tabularx}
\usepackage{graphicx}
\usepackage{fancyhdr}
\title{A Fault Tolerant, Area Efficient Architecture\\for Shor's Factoring Algorithm\\[-0.35in]}

\numberofauthors{1}
\author{
\alignauthor
Mark G. Whitney, Nemanja Isailovic, Yatish Patel and John Kubiatowicz\\
\affaddr{Computer Science Division}\\
\affaddr{University of California, Berkeley}\\
\email{\{whitney, nemanja, yatish, kubitron\}@eecs.berkeley.edu}
}

\begin{document}

%% Bullets in itemize closer together
\renewcommand{\itemhook}{\setlength{\itemsep}{0pt}}

%% Don't place large figures alone on a page
\renewcommand{\floatpagefraction}{0.8}

%% Macros for use
\newcommand{\etal}{\emph{et.al.}}
\newcommand{\eg}{\emph{e.g.}}
\newcommand{\ie}{\emph{i.e.}}

\maketitle

\vspace*{-0.5in}
\begin{abstract}

We optimize the area and latency of Shor's factoring while
simultaneously improving fault tolerance through: (1) balancing the
use of ancilla generators, (2) aggressive optimization of error
correction, and (3) tuning the core adder circuits.  Our custom CAD
flow produces detailed layouts of the physical components and utilizes
simulation to analyze circuits in terms of area, latency, and success
probability.  We introduce a metric, called ADCR, which is the
probabilistic equivalent of the classic Area-Delay product.  Our error
correction optimization can reduce ADCR by an order of magnitude or more.
Contrary to conventional wisdom, we show that the area of an optimized
quantum circuit is \emph{not} dominated exclusively by error
correction.  Further, our adder evaluation shows that quantum
carry-lookahead adders (QCLA) beat ripple-carry adders in ADCR,
despite being larger and more complex.  We conclude with what we
believe is one of most accurate estimates of the area and latency
required for 1024-bit Shor's factorization: 7659 mm$^{2}$ for the
smallest circuit and $6 \times 10^8$ seconds for the fastest circuit.

% Utilizing a custom computer aided design flow, we produce
% area-efficient microarchitectural designs for a 1024-bit quantum adder
% circuit in ion trap technology.  Compared to prior designs for the
% same circuits, our architectures achieve superior levels of fault
% tolerance, lower latencies and much smaller areas.  We accomplish this
% by using a number of optimizations including serial adder
% architectures, selective placement of error recovery operations, and
% hardware specialization into memory, functional unit, and ancilla
% qubit production regions.  Using lessons from our adder design, we
% produce a physical design for an implementation of Shor's
% factorization algorithm.  Although not yet capable of handling
% 1024-bit factorization circuits, our CAD flow shows an improvement of
% 1.6 in latency and factor of 5 in area over state of the art
% techniques for 16-bit factorization; we anticipate that we will be
% able to handle much larger factorization circuits in the near
% future. Our CAD flow produces detailed layouts of the physical
% components and orderings for classical scheduling and utilizes
% simulation to analyze circuits in terms of latency and error
% propagation.

\end{abstract}

% LocalWords:  Shor's retiming ancilla

\renewcommand{\paragraph}[1]{\vskip 6pt\par%
\noindent{\bf #1}}
\sloppypar
\section{Introduction}\label{sec:intro}

Quantum computing shows great potential to speed up difficult
applications such as factorization~\cite{shor1995srd} and quantum
mechanical simulation~\cite{zalka1998sqs}.  Unfortunately, recent
attempts to estimate the area required for such important applications
have resulted in impractically large areas; for example, a large area
dedicated to fault tolerance led to one estimated chip size on the
order of $1m^2$\cite{metodi2005qla} for a 1024-bit Shor's factoring
algorithm. In this paper, we show that these large areas result from
an assumption that the underlying quantum data path is not
specialized---essentially a uniform ``sea-of-gates.''

Since Shor's factorization is such an important application of quantum
computing, we believe that it justifies significant effort to produce a
practical, optimized circuit.  This paper examines \emph{specialized}
circuits and layouts for Shor's factorization, using an ASIC-like tool
flow.  
%To optimize large quantum circuits, we must be particularly
%careful about errors introduced through data movement and quantum
%operations.  
Because error correction can easily dominate the area and latency of any
quantum circuit, we must avoid excessive error correction operations by
limiting movement, balancing resource usage, and selectively correcting
for errors.  Thus, we introduce a new composite metric for probabilistic
circuits, called \emph{Area-Delay-to-Correct-Result} (ADCR), which is a
quantum equivalent of the classic Area-Delay product.  ADCR permits
quick comparisons of the efficiency to which quantum layouts make use of
area.  One surprising insight that results from careful accounting is
that area devoted to error correction does \emph{not} dominate the area
of an optimized quantum circuit.

\begin{figure}
  \vspace*{-0.05in}
  \begin{center}
    \includegraphics[width=\columnwidth]{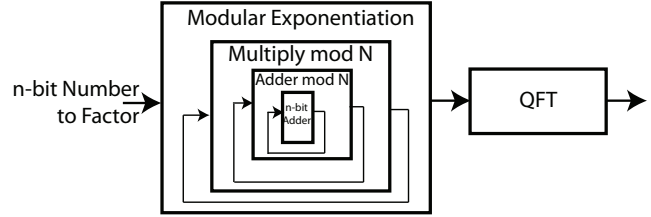}
  \end{center}
  \vspace*{-0.2in}
  \caption{Shor's Factoring Algorithm: The majority of the work in
    Shor's factoring algorithm for a n bit number is modular
    exponentiation.  At its core is repeated applications of quantum
    n-bit addition.}
  \label{fig:shor_arch_simple}
  \vspace*{-0.1in}
\end{figure}

In tackling Shor's factorization, we start with the structure of Shor's
algorithm, shown in Figure~\ref{fig:shor_arch_simple}.  Here, we see
that quantum addition is fundamentally at the core, suggesting that an
optimized adder circuit is important to efficient factoring.
Consequently, we take a three-pronged approach:

\paragraph{Balanced Ancilla Generation:}  We utilize
  shared ancilla factories~\cite{steane1999stp,isailovic2008rqc} to
  greatly reduce the area required for ancilla generation.
  Section~\ref{sec:mapping} presents
  mapping heuristics that allow us map a quantum circuit to compute
  regions with shared ancilla generators while automatically segregating
  idle data bits to special memory regions with lower ancilla bandwidth.
  It finishes by provisioning a custom teleportation network to
  communicate over longer distances.  

\paragraph{Aggressive Error Correction Optimization:} Since error
  rates in quantum computers are so high, most quantum computer
  architects have opted for a brute-force error correction strategy that
  corrects after every operation or movement.  Section~\ref{sec:qec}
  shows how a more intelligent placement of error correction operations,
  modeled after circuit retiming, can reduce ADCR by an order of magnitude.

%The rigorously
%  defined \emph{threshold theorem} \cite{aharonov1997ftq} says that if
%  we use fault tolerant quantum error correction (QEC) procedures
%  frequently in our circuit, we can operate at a constant logical data
%  error rate.  This is desirable but this and successive works do not
%  prescribe how often we must apply these QEC procedures to get our
%  desired data error rate.  Since different circuit topologies lead to
%  different error propagation patterns, this problem is non-trivial
%  and previous schemes opted for a brute-force, correct everywhere
%  strategy.  We show why this is a bad idea and how more intelligent
%  QEC can lead to better fault tolerance and resource utilization.

\paragraph{Tuning the Parallelism of Adder Circuits:} Since the quantum
  adder circuit \cite{draper2004ldq,vedral1996qne} is \emph{the} core
  component to Shor's factoring algorithm, Section~\ref{sec:adderdesign}
  investigates microarchitectures and layouts for quantum adders.  We
  believe that this is the first work to examine large quantum adder
  architectures in such detail.

\vspace*{12pt}\noindent At the core of this work is a custom computer
aided design (CAD) flow for synthesizing fault-tolerant quantum layouts.
We can accurately evaluate the area, latency and
fault tolerance of circuits\footnote{Note that some of the partitioning heuristics
that we describe in Section~\ref{sec:mapping} are required to
\emph{automatically} evaluate organizations with specialized memory
regions, such as CQLA~\cite{thaker2006qmh}; previous evaluations were
hand partitioned.}.  We can directly compare the efficiency of various
quantum datapath organizations; Section~\ref{sec:archs} details these
organizations and our evaluation methodologies.  
 
% Ultimately, we produce a complete, optimized implementation of Shor's
% algorithm for factoring.  
Our investigation of Shor's algorithm is very detailed (see
Figure~\ref{fig:ShorComplete} in Section~\ref{sec:shor}).  Although it
is hard to directly compare with existing proposals which contain many
estimates (such as \cite{metodi2005qla, thaker2006qmh}), we are
confident that (1) our methodology provides one of the most realistic
evaluations of area and latency for 1024-bit Shor's factoring and (2)
our layouts are an important step forward toward optimal factoring
circuits.

% The rest of the paper is as follows: Section~\ref{sec:archs} presents
% basic quantum computer organizations that we explore and discusses
% how we evaluate our results.  Section \ref{sec:mapping} describes
% our heuristics for partitioning and resource distribution.  Section
% \ref{sec:qec} presents our quantum error correction optimization.
% Section \ref{sec:adderdesign} explores quantum adder designs. Finally,
% in Section \ref{sec:shor}, we present our full design for
% factorization. Section~\ref{sec:related} discusses related work, and
% Section \ref{sec:conclusion} concludes.

% LocalWords:  Shor's Ancilla ancilla retiming microarchitecture CQLA

\pagestyle{empty}
\section{Architectural Exploration}\label{sec:archs}

This section describes the quantum datapath organizations that we will
use in other sections.  It also introduces our CAD flow,
error analysis, and evaluation metrics.

\subsection{Abstracting Ion Trap Computers}\label{sec:abstract}

\begin{figure}
%\vspace{0.1in}
\begin{center}
\includegraphics[width=\columnwidth]{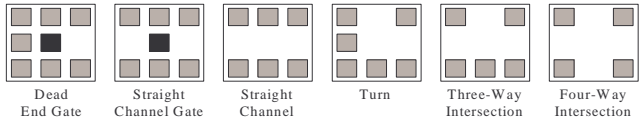}
\end{center}
\vspace{-0.25in}
\caption{Macroblocks: Abstract building blocks for Ion Trap layouts.
Black boxes are gate locations, gray boxes are abstract ``electrodes,''
and wide white channels are valid paths for qubit movement}
\label{fig:macroblocks}
\vspace{-0.12in}
\end{figure}

We utilize Ion Trap technology~\cite{seidelin06} as the substrate for
quantum datapaths. Table~\ref{table:ion_trap_basic_stats} shows basic
error rates and latencies. It includes two ``Error Sets,'' one that we
believe represents state of the art (Set 1) and one with a higher fault
rate to explore error correction (Set 2). As described
elsewhere~\cite{isailovic2008rqc,whitney2007agl}, our layout and
simulation infrastructure utilizes an abstraction of Ion Traps,
mentioned briefly here:

\begin{itemize}
\item{\bf Qubits:} 
Qubits (ions) are represented by their position in the
substrate as well as fidelity (level of error).

\item{\bf Layout:} We utilize the \emph{macroblock} abstraction, shown
in Figure~\ref{fig:macroblocks}.  Each macroblock has one or more
``ports'' through which qubits may enter and exit and which connect to
adjacent macroblocks.  To perform a gate operation, involved qubits must
enter a valid gate location and remain there for the duration of the
gate.  A layout for a circuit consists of a tiling of macroblocks and a
schedule for qubit movement and operations.

\item{\bf Movement:} Trapped ions are moved via pulses applied to 
electrodes.  Since ions are trapped, they can only
move along channels specified by the layout.

\item{\bf Gates:} Gates are performed by firing laser pulses at trapped
ions.  We abstract the physics and consider that a gate is performed after
arrival of appropriate qubits at ``gate locations'' in the
layout.
\end{itemize}
This representation provides sufficient accuracy to evaluate the area,
latency, and error behavior of quantum circuits.

\begin{table}
\begin{center}
{\small
\begin{tabular}{|l|c|c|c|}
\hline
		   & Error & Error & Latency\\
Physical Operation & Set 1~\cite{MEMStrap04madsen} & Set 2~\cite{steane2004bbg} & in ($\mu$s)~\cite{ozeri95hcp} \\
\hline \hline
One-Qubit Gate & $10^{-6}$ & $10^{-4}$ & 1 \\
Two-Qubit Gate & $10^{-6}$ & $10^{-4}$ & 10 \\
Measurement    & $10^{-6}$ & $10^{-4}$ & 50 \\
Zero Prepare   & $10^{-6}$ & $10^{-4}$ & 51 \\
\hline
Straight Move ($\sim$30 $\mu$m) & $10^{-8}$ & $10^{-6}$ & 1 \\
90 Degree Turn & $10^{-8}$ & $10^{-6}$ & 10 \\
\hline
Idle (per $\mu$s) & $10^{-10}$ & $10^{-8}$ & N/A \\
\hline
\end{tabular}}
\end{center}
\vspace{-0.2in}
\caption{Error probabilities and latency values used by our CAD flow
for basic physical operations
%~\cite{steane2004bbg,isailovic2008rqc}.
}
\label{table:ion_trap_basic_stats}
\end{table}

\begin{figure*}[t!]
\begin{minipage}{\hsize}
\begin{center}
\includegraphics[width=\textwidth]{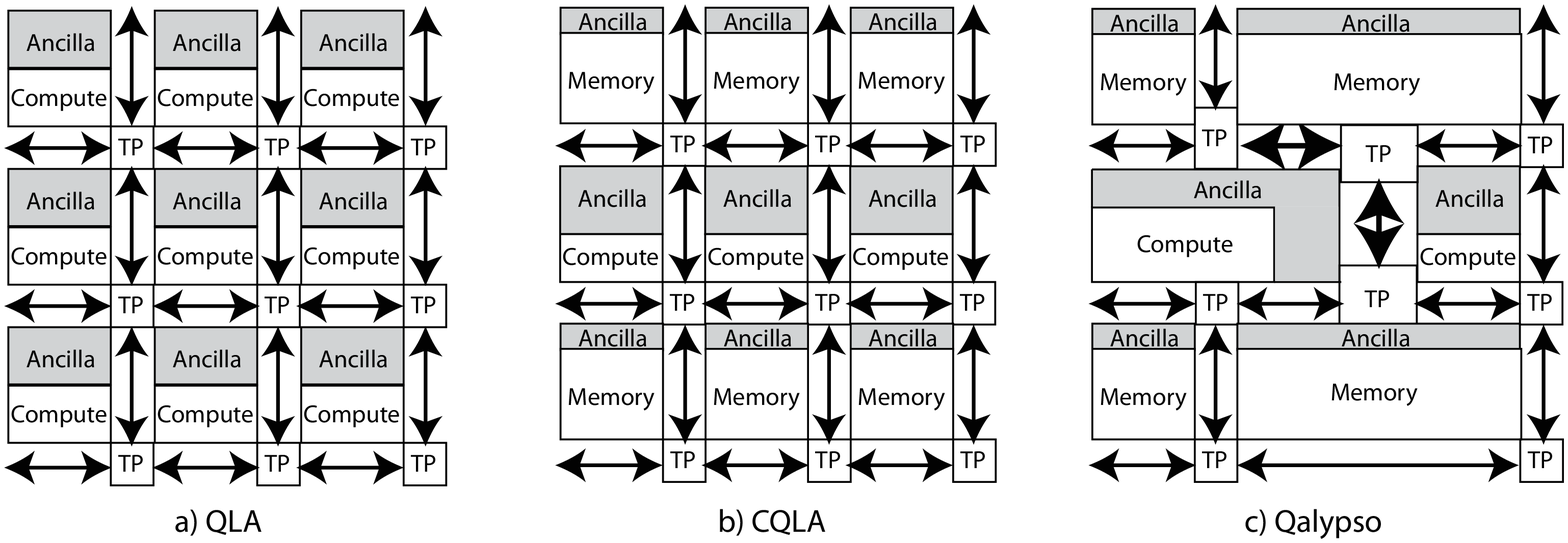}
%,width=0.85\hsize}
\end{center}
\vspace{-0.27in}
\caption{Quantum Datapath Organizations: a) Quantum Logic Array
  (\emph{QLA}): An FPGA-style sea of quantum two-bit gates (compute
  tiles), where each gate has dedicated ancilla resources.  b)
  Compressed QLA (\emph{CQLA}): QLA compute tiles surrounded by denser
  memory tiles. c) ~\emph{Qalypso}: Variable sized compute and memory
  tiles with shared ancilla resources for each tile;
  teleportation network can have variable bandwidth links.}
\vspace{0.12in}
\label{fig:cqla}
\end{minipage}
\begin{minipage}{\columnwidth}
\begin{center}
\includegraphics[width=\textwidth]{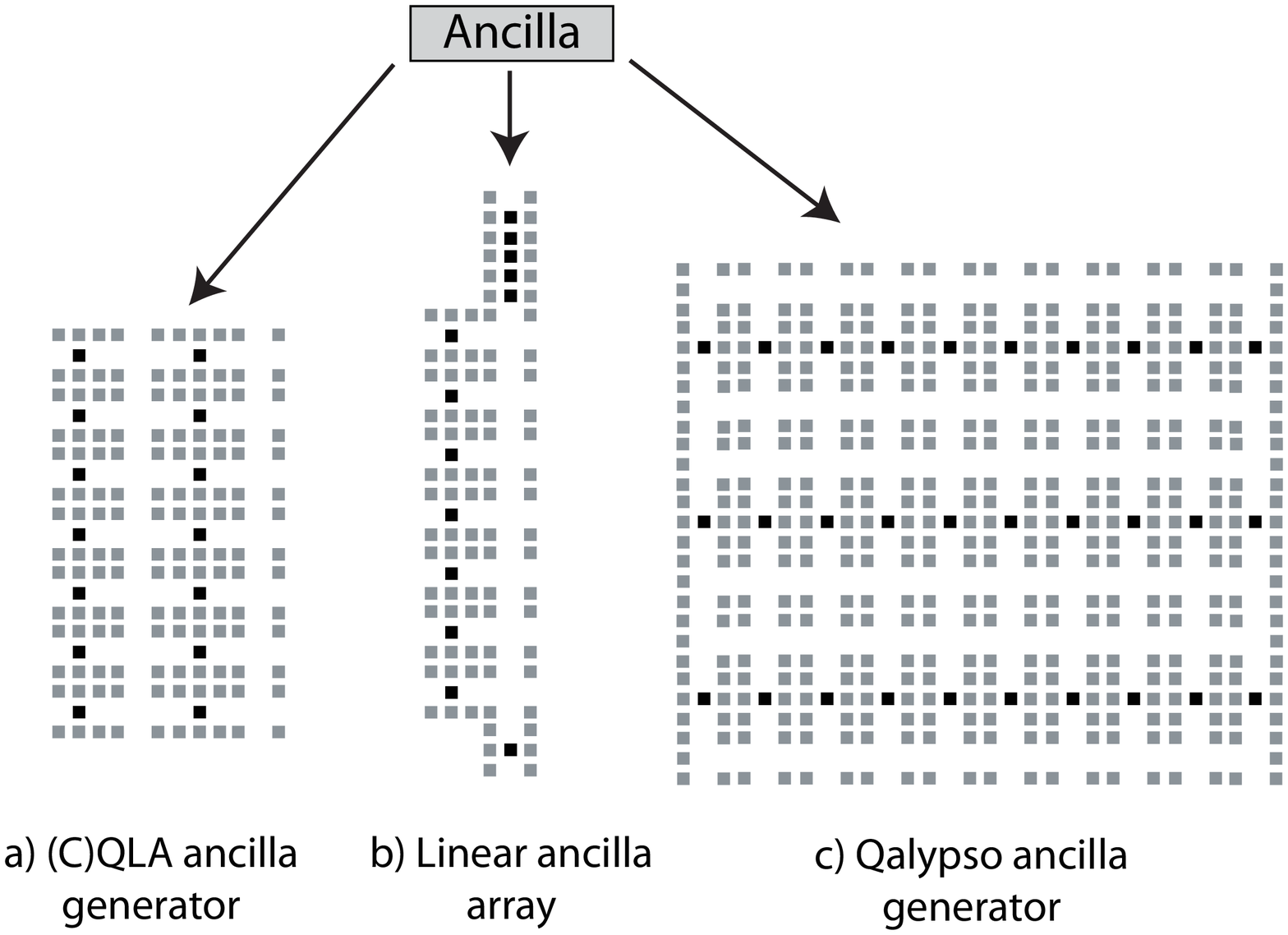}
%,width=0.95\hsize}
\end{center}
\end{minipage}\hfill
\begin{minipage}{\columnwidth}
  \renewcommand {\tabularxcolumn}[1]{>{\arraybackslash}m{#1}}
  \begin{tabularx}{0.95\hsize}{|c|X|}
    \hline
    \small Datapath & Description\\
    \hline\hline
    QLA & Original Quantum Logic Array \cite{metodi2005qla}, compute
    regions only, no specialization\\
    \hline
    LQLA & QLA with an optimized ancilla generator from
    \cite{kregerstickles2008mai}\\
    \hline
    CQLA & Compressed QLA \cite{thaker2006qmh}, compute and memory
    regions specialization, original ancilla generator\\
    \hline
    CQLA+ & CQLA with a better performing ancilla generator
    from \cite{steane2003oan}\\
    \hline
    Qalypso & Our architecture \cite{isailovic2008rqc}.  Variable
    sized compute and memory regions, variable resources in ancilla
    generators and teleport network.  ``Pipelined'' ancilla factory
    optimized from design in \cite{steane2003oan}.\\
    \hline
  \end{tabularx}
\end{minipage}\vspace*{-0.05in}
\begin{minipage}[t]{\columnwidth}
\caption{Ancilla generation unit layouts from 3 different
  architectures.}
\label{fig:ancilla_factories}
\end{minipage}\hfill
\begin{minipage}[t]{\columnwidth}
\caption{Taxonomy of the quantum computer datapath organizations we
    investigate in this work.}\label{table:arch_list}
\end{minipage}
\vspace*{-0.05in}
\end{figure*}

\subsection{Quantum Datapath Organization}\label{sec:datapath}

Proposed architectures for quantum computers have all consisted of
computation regions connected by an interconnection network using
quantum teleportation \cite{metodi2005qla, isailovic2006ins}.  High
fault rates in quantum computing necessitate the widespread use of
quantum error correction (QEC).  Further, ancilla state
generation is important to aid in the correction process
\cite{steane1999stp} and as an integral part of quantum algorithms.

\paragraph{Three Major Organizations:} Figure~\ref{fig:cqla} shows three
major \emph{datapath organizations} that represent the ``state of the
art'' in quantum computing\footnote{Since circuits are \emph{mapped} to
  these datapaths, they are not quite ``architectures'' but rather raw
  material for constructing architectures.}.  
% We refer to these as ``datapath
% organizations'' instead of quantum computing architectures, since the
% architecture of a quantum computing circuit consists of higher-level
% structures that are ultimately mapped to an underlying datapath
%organization.  
They are QLA~\cite{metodi2005qla}, CQLA~\cite{thaker2006qmh}, and
Qalypso~\cite{isailovic2008rqc} and can be viewed as a spectrum from
inflexible to flexible ancilla distribution.  They differ in their
configuration of compute regions, ancilla generation areas, memory
regions (for idle qubits), and teleportation network resources (for
longer-distance communication)\cite{isailovic2006ins, metodi2005qla}.

The QLA architecture is most like a classical FPGA, in that all elements
are identical: each element contains enough resources to perform a
two-bit quantum gate.  Each such \emph{compute region} contain dedicated
ancilla generation resources, space for two encoded quantum bits, and a
dedicated teleportation router for communication.

CQLA improves upon QLA by allowing two different types of data regions:
\emph{compute regions} (identical to those in QLA) and \emph{memory
regions} (which store eight quantum bits)~\cite{thaker2006qmh}.  To
account for different failure modes (idle errors vs interaction errors),
data in memory regions are encoded differently from data in compute
regions.

Finally, Qalypso improves upon CQLA by further relaxing the strict
assignment of ancilla generation resources.  It allows optimized,
pipelined ancilla generators to feed regions of data bits (compute
regions) that can perform more than just two-bit gates.  The sizing of
ancilla generators and data regions can be customized based on circuit
requirements.  Qalypso requires analysis (Section~\ref{sec:mapping}) to
balance ancilla consumption with ancilla generation.  Such analysis can
automatically adjust the amount of ancilla bandwidth required in
memory regions based on the residency time of qubits.

In all three organizations, each compute or memory region is placed adjacent
to a teleport router.  Qubits are moved ballistically within regions and
teleported between regions.

\paragraph{Custom Component Design:}
Proper design of the datapath elements (such 
teleportation routers or ancilla generators) is an important factor.  
%Incorrect design could
%lead to a mapped circuit that is too large, fails frequently, or
%performs poorly.  
In this paper, we pay careful attention to the teleportation
network~\cite{isailovic2006ins, metodi2005qla}.  We have produced
layouts for the routers and EPR generators and utilize these in
computing area, latency, and error probability of circuits.
Sections~\ref{sec:errors} and~\ref{sec:networkmap} discusses how these
numbers are derived and integrated with our evaluation methodology.

We also investigate a number of options for ancilla generation, as shown
in Figure~\ref{fig:ancilla_factories}.  The original QLA and CQLA papers
used the ancilla generator shown in Figure~\ref{fig:ancilla_factories}a.
Recent work by Kreger-Stickles and Oskin~\cite{kregerstickles2008mai}
%investigates layout and scheduling of ancilla factories for 7-bit error
%correction codes; it 
showed that a layout such as
Figure~\ref{fig:ancilla_factories}b, which is designed to minimize qubit
movement and idling, is more fault tolerant and faster than the original
QLA/CQLA ancilla factory under certain fault assumptions.  We will insert
this generator into the QLA datapath and refer to the result as LQLA.
Finally, Figure~\ref{fig:ancilla_factories}c shows a more fault-tolerant
variant of the original QLA/CQLA ancilla generator used in
Qalypso~\cite{isailovic2008rqc}.  It performs three copies of
the QLA/CQLA generation process to produce a higher fidelity result.

\begin{figure*}
  \begin{center}
    \includegraphics[width=\textwidth]{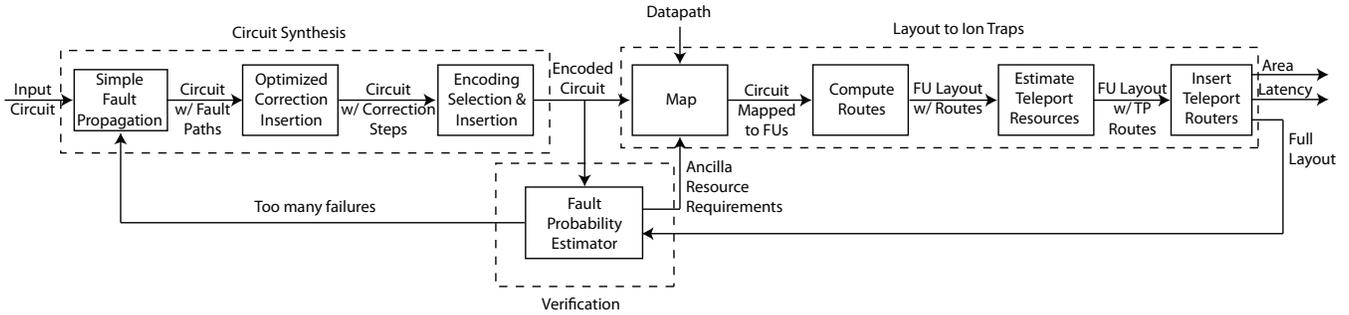}
%,width=0.95\hsize}
  \end{center}
\vspace{-0.3in}
  \caption{At the core of our microarchitectural evaluation is this
    custom computer aided design flow.  It consists of 3 segments,
    circuit synthesis \& optimization, layout, and
    verification.}\label{fig:methodology_flow}
\vspace{-0.1in}
\end{figure*}

\paragraph{An Organizational Zoo:} Figure~\ref{table:arch_list} shows the
datapath organizations that we will use in this paper.  Three of these,
namely QLA, CQLA, and Qalypso come directly from their original papers.
LQLA is a variant of QLA utilizing the new ancilla generator and cell
layout proposed by Kreger-Stickles.  Finally, CQLA+ is a version of CQLA
utilizing the improved ancilla generator from the Qalypso paper.

Evaluating such a disparate set of architectures is always challenging.
When possible, we have adapted the exact qubit scheduling provided by
authors (such as in LQLA, where the authors provided us with scheduling
of qubits for their ancilla generator).  To evaluate larger circuits, we
have developed a hybrid evaluation methodology (described in
Section~\ref{sec:errors}) that permits us to stitch together modules.

We stray as little as possible from published organizations, with a few
exceptions.  First, our emulation of CQLA does not use a
different ECC code between memory and compute regions.  Second,
assignment of qubits to memory regions happens automatically according
to the mapping heuristics of Section~\ref{sec:mapping}, rather than
through hand-partitioning.  Third, LQLA is our invention,
produced by inserting the optimal ancilla factory
from~\cite{kregerstickles2008mai} into QLA; this was necessary because
the authors of~\cite{kregerstickles2008mai} did not take a stand on
long-distance communication or memory regions.  
   
\subsection{Synthesis of Application Specific Circuits}\label{sec:synthesis}

% Quantum programs are initially specified as mathematical equations.
% In~\cite{eval_framework}, Balensiefer et al introduced a quantum
% assembly language (QASM) more suited to execution on a quantum
% machine.  QASM specifies a quantum program in terms of fundamental
% physical operations, in addition to classical control operations and
% predicates.  The quantum operations in QASM correspond directly to
% atomic operations on the datapath, and thus a classical analogue of
% QASM would be a program specification in terms of NANDs, NORs and
% NOTs.
% 
% Implementing a given application QASM file on a quantum computer
% involves mapping the QASM instructions to a set of computation units
% in a given substrate.  Quantum datapath designs generally focus on
% tiled general-purpose compute elements to execute these QASM programs.
% Due to the fragility of quantum state, each ``logical qubit'' (a qubit
% used in a quantum program) must be encoded into several ``physical
% qubits'' (components on the datapath).  Logical qubits must
% periodically undergo quantum error correction (QEC)
% ~\cite{knill1997tqe, shor1995srd, steane1996ecc}, so each tile must
% somehow allow for hardware to perform QEC (discussed further in
% Section~\ref{sec:mapping}).
% 

We have implemented a computer aided design (CAD) flow for quantum
circuits.  As shown in Figure \ref{fig:methodology_flow}, it consists of
3 main pieces: synthesis and optimization of error correction, mapping
and layout of gates onto a substrate, and verification of fault
tolerance properties.  

Quantum circuits are specified in QASM, a quantum assembly language
originally developed by Balensiefer \emph{et. al}~\cite{eval_framework}.
We have extended QASM with primitives that allow specification of
control operations and better hierarchical design.  One view of QASM is
that it is a quantum netlist format, permitting the description of
quantum circuits as a set of interconnected gates and scheduling
constraints.  Although some of the details of our CAD flow are prosaic,
this flow is considerably more detailed, flexible, and complete than
reported by other authors.

To map a QASM specification to one of the datapath organizations from
Figure~\ref{table:arch_list}, we must map gate sequences to the compute
and memory regions.  Further, we determine the amount of ancilla
necessary to perform error correction at each region and provision
teleportation network resources to provide necessary bandwidth.  Details
of the mapping heuristics are presented in Section \ref{sec:mapping}.

The error control mapping phase is one of the innovations of this paper.
It takes a logical circuit, produces a redundantly encoded circuit in a
specified quantum ECC code, then selectively adds error correction
operations.  We have developed a novel optimization procedure for
slashing error correction overhead by a factor of three (3) over
existing techniques from the literature, as discussed in Section~\ref{sec:qec}.

\subsection{Assessing the Error of Quantum
  Circuits}\label{sec:errors} 
%As mentioned in
%Section~\ref{sec:synthesis}, our synthesis flow automatically encodes
%quantum circuits in an error correction code of choice and inserts
%periodic error correction operations.  
After a circuit has been mapped to one of the datapaths described in
Section~\ref{sec:datapath}, we must evaluate the resulting error
behavior.
% in order to compare the advantages and disadvantages of
%different datapaths.  
Circuits are typically large and hierarchically
specified, leading us to develop a \emph{hybrid error modeling}
technique, described in this section.

%Consistent with a variety of other works \cite{eval_framework,
%localfault04svore}, we break down errors into three categories:
As usual, we classify errors into three categories~\cite{eval_framework,
localfault04svore}:
\begin{itemize}
\item{\bf Gate:} Errors that occur while manipulating quantum
  data. This category is the most widely
  studied in association with error correction performance.  
% Much of
%  the work in fault tolerance is to assure that the errors from the
%  gates in an error correction procedure do not corrupt a data qubit
%  more than it helps in recovering from previous errors
%  \cite{aharonov1997ftq, preskill1997ftq, aliferis2005qat}.  
  It is generally believed that gate errors will be the most significant of
  the three types.
%, since it involves deliberate manipulation of quantum
%  state.
\item{\bf Movement:} Errors that result from moving quantum data.  
%Movement of quantum data is generally error prone.
  For example, ballistic ion movement in trapped ion technology
  involves accelerating and decelerating ions both linearly and around
  corners\cite{hensinger2006tji}.  This induces unwanted
  vibrations or even collisions, disrupting the internal state that
  represents the datum.
\item{\bf Idle:} Errors in idle qubits that result from spontaneous
  quantum effects.  It is generally thought that idle or storage errors
  are the least severe of the three per unit time for most technologies.
\end{itemize}
In principle, we must simulate a circuit from start to finish, injecting
errors by assuming that every gate, qubit movement, and qubit stall has
an associated error rate and that every error event injects either a bit
flip, phase flip or both.  Unfortunately, interesting circuits are too
large to do this exactly---leading to a need for a hybrid methodology.

% Variables we use:
%
% Number of Data regions in a datapath
\def\ndata{D}
% Number of Qubits in a Data region
\def\ndataq{D_{q}}
% Number of Ancilla Generators in a Data region
\def\ndataa{D_{ag}}

% Number of Memory regions in a datapath
\def\nmem{M}
% Number of Qubits in a Memory region
\def\nmemq{M_{q}}
% Number of Ancilla Generators in a Memory region
\def\nmema{M_{ag}}

\begin{table*}
  \renewcommand {\tabularxcolumn}[1]{>{\arraybackslash}m{#1}}
\begin{tabularx}{\hsize}{| c || c | c | c || c | c | c || c | >{\hsize=0.75\hsize\centering\arraybackslash}X | >{\hsize=1.25\hsize\raggedright\arraybackslash}X |}
\hline
         & \multicolumn{3}{c||}{Data Regions} & \multicolumn{3}{c||}{Memory Regions} & Ancilla & \multicolumn{1}{c|}{Non-Trans} & \\ \cline{2-7}
Datapath & Total & Qubits & Gens & Total & Qubits & Gens & Generator & \multicolumn{1}{c|}{Gates} & \multicolumn{1}{c|}{Interconnect}\\
\hline \hline
QLA & $\ndata$ &  2 & 2 & \multicolumn{3}{c||}{none} & \cite{metodi2005qla} & anywhere & fixed-size routers, one per data/memory region \\
\hline
LQLA & $\ndata$ & 2 & 2 & \multicolumn{3}{c||}{none} & \cite{kregerstickles2008mai} & anywhere & fixed-size routers, one per data/memory region \\
\hline
CQLA & $\ndata$ & 36  & 36 & $\nmem$ & 64 & 8 & \cite{metodi2005qla,thaker2006qmh} & anywhere & fixed-size routers, one per data/memory region \\
\hline
CQLA+ & $\ndata$ & 36 & 36 & $\nmem$ & 96 & 12 & \cite{isailovic2008rqc} & anywhere & fixed-size routers, one per data/memory region \\
\hline
Qalypso & $\ndata$ & $\ndataq$ & $\ndataa$ & $\nmem$ & $\nmemq$ &
         $\nmema$ & \cite{isailovic2008rqc} & placed with custom
         ancilla & variable sized routers adapted to design \\
\hline
\end{tabularx}
\vspace{-0.15in}
\caption{Details of various datapath organizations.  A datapath consists
of a number of Data Regions and in some cases Memory Regions.  Each
Data/Memory region is sized to hold a specific number of Qubits and
Ancilla Generators (Gens) and regions are connected via an
Interconnection Network.  The $D$ and $M$ variables in the table signify
values that are only determined after a quantum circuit is mapped onto
the datapath.}
\vspace{-0.05in}
\label{table:arch_details}
\end{table*}

\paragraph{Hybrid Error Modeling:} Large designs are specified
hierarchically, as a tree of modules.  While we can synthesize a
complete macroblock layout with fine-grained placement and routing for
smaller modules, high-level modules are better handled via
coarse-grained mapping techniques.  Our mapper does not create exact
macroblock specifications for all inter-block channels but instead
relies on estimates of ballistic movement and teleportation based on
inter-block distances.  Further, the distance traveled from ancilla
factories to data bits is estimated (quite accurately) after data bits
have been placed.  Consequently, we utilize a hybrid simulation model in
computing communication costs and qubit idle times: not every qubit
movement is simulated, but rather aggregate movements are computed and
combined to speed simulation and support hierarchical design.

The calculation of the error probability for a mixed ballistic movement
model involves three types of information: exact error probabilities,
errors from estimated ballistic channels, and errors from teleportation
channels.  For smaller, leaf modules, we extract error properties
exactly through simulation.  The coarse-mapped distance estimates for
longer ballistic communications are translated into a count of straight
and turn macroblocks traversed, yielding error fidelity numbers for
traversing these channels.

%\begin{figure}
%\begin{center}
%\epsfig{file=figures/teleport_movement_estimates,width=0.85\hsize}
%\end{center}
%\vspace{-0.2in}
%\caption{\label{fig:teleport_movement} Different components of the
%  inter-module communication network that we estimate in order to
%  perform accurate error simulation.}
%\vspace{-0.1in}
%\end{figure}

Finally, the effects of teleportation are determined by computing the
fidelity and bandwidth of EPR bits in the channel.
%Figure \ref{fig:teleport_movement} shows the different teleportation-involved
%inter-block movement and idle operations that we account for in our
%error simulations.  
We have a model of EPR generation, routing, and purification which
permits accurate computation of the latency to setup a teleportation
channel as well as the fidelity of the EPR bits available for it.  We
compute the gate and movement errors within routers along the path, EPR
generators, and purifiers.  Section~\ref{sec:networkmap} gives more
details.

Consequently, to compute the error probability of a large,
hierarchically specified circuit, we combine inter-block movement and
idle errors from ballistic and teleportation channels with the exact
gate, movement, and idle errors in the compute and ancilla regions to
produce one sequential list of possible error points in the layout as
the circuit executes.  This error list is passed to the Monte Carlo
error simulator.

\paragraph{Monte Carlo Error Simulation:}
To propagate errors, we utilize Monte Carlo (MC) simulation
\cite{steane2003oan, cross2007ccs}, in which errors are sampled for each
circuit element and errors are propagated accordingly.  We
traverse a graph representing the full circuit and layout, sampling
gate, movement and idle errors (some of which are aggregate error counts
for ballistic and teleportation-based communication) on each qubit in
dataflow order.  If the final state of the qubits results in an
uncorrectable error, the run is counted as a failure.  This process is
repeated many times to get a statistically significant sample.  Our tool
uses the Colt JET library~\cite{colt2004} version of the Mersenne Twister random number generator.

\subsection{Evaluating Designs}\label{sec:evaluation}

In the following sections, we evaluate datapath organizations and layout
heuristics via the following four metrics:

%\paragraph{Operation counts:} We can count the total number of gates needed
%  to perform a computation.  This is a constant as we trade off area
%  (providing additional parallelism) and latency.  In the context of
%  this paper, the number of \emph{logical} operations remains constant
%  and the operation count refers the to total number of
%  \emph{physical} operations and can vary only by the choice and
%  amount of error correction performed.  

\paragraph{Area:} To measure the \emph{area} of a
quantum circuit, we first \emph{map} that circuit to particular quantum
datapath, then count the resulting number of macroblocks (see Section~\ref{sec:abstract}).

\paragraph{Latency:} The \emph{latency} (Latency$_{\textrm{single}}$) of a quantum circuit 
represents the time to evaluate that circuit on a given layout.  After
mapping the circuit (including scheduling its operations), we measure
the latency through simulation.

\paragraph{Success Probability:} The \emph{success probability}
($P_{\textrm{success}}$) for a quantum circuit represents the probability that
the result will be correct (error free) after evaluation.  The success
probability is measured with hybrid simulation
(Section~\ref{sec:errors}).

\paragraph{Area-Delay-to-Correct-Result (ADCR):}  
%%Since we are
%%targeting efficient use of area to improve latency and raise success
%%probability, 
To evaluate the quality of quantum layouts, we propose a
\emph{composite} metric called \emph{Area-Delay-to-Correct-Result
(ADCR)}.  ADCR is the probabilistic equivalent of the Area-Delay
product:
%%\begin{equation}
%%    ADCR = Area\times E(Latency) = Area\times \frac{Latency}{P_{success}}
%%    ADCR = Area\times Latency\times\sum_{n=1}^{\infty}{P_{success}(1-P_{success})^{n-1}}
%%  Area\times Latency\times\sum_{n=1}^{\infty}{P_{success}(1-P_{success})^{n-1}} = Area\times \frac{Latency}{P_{success}}
%%\end{equation}
{\small\begin{align*}
   \textrm{ADCR} & = \textrm{Area}\times E(\textrm{Latency$_{\textrm{total}}$}) \\
      & = \textrm{Area}\times \sum_{n=1}^{\infty}{n\cdot\textrm{Latency$_{\textrm{single}}$}\cdot P_{\textrm{success}}(1-P_{\textrm{success}})^{n-1}} \\
      & = \textrm{Area}\times \frac{\textrm{Latency$_{\textrm{single}}$}}{P_{\textrm{success}}} 
\end{align*}}%
For ADCR, \emph{lower is better}.  By incorporating potential for
circuit failure, ADCR provides a useful metric to evaluate the area
efficiency of probabilistic
circuits.  It highlights, for instance, layouts that use less area for
the same latency and success probability.  Or, layouts that use the same
area for lower latency or higher success probability.
 
%%ADCR highlights, for instance,
%%layouts that use less area for the same latency and success probability.
%%Or, layouts that use the same area for lower latency or higher success
%%probability.  ADCR provides a useful metric to evaluate probabilistic
%%circuits, since we are ultimately interested in the expected
%%time to a correct result.
%%: single-pass latency scaled by probability of
%%success.

% LocalWords:  datapath datapaths Qubits macroblock qubits macroblocks qubit et
% LocalWords:  QLA FPGA ancilla CQLA Qalypso teleportation LQLA teleport QEC al
% LocalWords:  Pipelined pipelined balistically teleported Kreger Stickles EPR
% LocalWords:  Oskin microarchitectural ECC QASM Balensiefer netlist dataflow
% LocalWords:  teleporting Mersenne cx ADCR

\section{Mapping Circuits to Datapaths}\label{sec:mapping}

%% The datapath organizations from Section~\ref{sec:datapath} provide us with
%% general structures onto which we will map specific quantum circuits.  
In this section, we discuss the process of mapping a quantum circuit to
one of the datapath organizations from Section~\ref{sec:datapath}.  
%The
%mapper customizes any adjustable parameters of the datapath while
%deciding \emph{when} and \emph{how} datapath resources will be used.
Table~\ref{table:arch_details} details the various parameters that must
be determined for each datapath organization.  Many of these parameters
are derived directly from papers by the various authors; only Qalypso
provides complete flexibility in the number of Qubits ($D_{q}$) and
Ancilla Generators ($D_{ag}$) per data region as well as number of
Qubits ($M_{q}$) and Ancilla Generators ($M_{ag}$) per memory region.

%\begin{figure}
%\begin{center}
%\epsfig{file=figures/annotated_CQLA.eps,width=0.9\hsize}
%\end{center}
%\vspace{-0.2in}
%\caption{}
%\label{fig:annotated_CQLA}
%\end{figure}

%% Since many interesting quantum circuits are too large to effectively map
%% directly onto a flat datapath, we hierarchically annotate our datapath
%% prior to mapping.  Our CAD flow analyzes each submodule to determine
%% available parallelism and decide on a set of compute and memory
%% resources needed to execute it.  The flow then identifies regions of the
%% target datapath which are suitable for each submodule.

\subsection{Partitioning the Circuit}

%% \subsection{Initial Datapath Sizing}
During the mapping process, the mapper must determine the total number
of data regions ($\ndata$) and memory regions ($\nmem$) required.  For
QLA and LQLA, $\nmem=0$ which makes it easy to determine $\ndata$ as
it is simply sized to accommodate the number of qubits used in the
quantum circuit.  The addition of memory regions introduces trade-offs
in area and exploitable parallelism (latency).  A datapath with a
single compute region and a sea of memory can only perform one
operation at a time --- resulting in longer latency with a minimal
area.  A datapath such as QLA with all compute regions and no memory
can exploit all possible parallelism in the circuit but with extremely
high cost in area.  

%% We perform automated analysis on the quanutm circuit to calculate
%% maximum, average, and standard deviation usage of each operation type
%% used. Then we use this to determine an appropriate set of functional
%% units needed to execute the circuit.  For example, we may find that the
%% 4-bit Carry Lookahead Adder uses two different operations: the Hadamard
%% (basic) operation with an average of 0.15 gates per time unit and
%% standard deviation of 0.13; and the Toffoli module with an average of
%% 1.93 uses per time unit and standard deviation of 0.5.  This particular
%% pattern results from the fact that each qubit in the Adder module
%% undergoes a Hadamard at the beginning and then undergoes some series of
%% Toffoli module.  Based on this, we may decide to use a datapath which
%% includes two Toffoli functional units, one or two Hadamard functional
%% units and a minimal of memory cells.

%% Our automated method results in a reasonable estimate for the number
%% of compute regions and memory regions needed to run a given circuit
%% while minimizing [FIXME some area/latency metric]. Alternatively, we
%% can manually specify a datapath configuration for evaluation. 

%% Once we are provided with a Datapath specification we move on to
%% mapping the quantum circuit onto the Datapath.  
The mapper determines where each data qubit will reside during
the course of the execution as well as when and where each quantum gate
will execute.  It starts with a coarse-grained partitioning of modules
to compute-regions that minimizes communication.  Next, the mapper
attempts to schedule each gate operation so that it occurs as late as
possible, while prioritizing operations on the critical path.  The
mapper relocates qubits into memory regions (if available) to free up
compute regions for subsequent operations.  As the mapper progresses, it
tracks the location and times of all gate operations, error corrections,
and network connections needed to perform the quantum circuit.  The
mapper discourages imbalanced mappings, such as those that over utilize
network links or ancilla generation resources.

If the target datapath has fixed ancilla generation resources, the
mapper attempts to map operations to regions with unused ancilla bandwidth.
%% If no resources are available, operations must wait
%% until resources are available.  
In datapaths with flexible ancilla generation (\eg, Qalypso),
the mapper assumes that operations will never wait for ancilla, while
still attempting to balance ancilla usage.  A later phase (described
below) matches ancilla generation resources to demand.

\begin{figure}
 \includegraphics[width=\columnwidth]{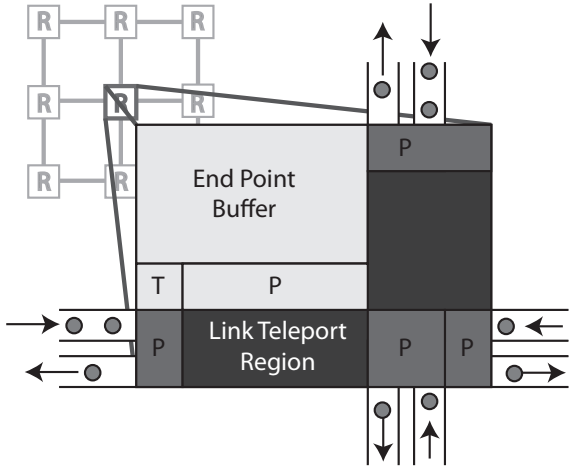}
%, width=0.78\hsize}
 \vspace{-0.15in}
 \caption{Network Router. Dark gray areas support single hop
links between neighboring routers.  Light gray regions handle
connections that terminate locally.  The size of the End Point Buffer is
dictated by the size of the logical qubit being teleported.}
 \label{fig:router_layout}
 \vspace*{-0.15in}
\end{figure}

\subsection{Ancilla Resources for Qalypso}
Ancilla generation for four of the datapath organizations (\emph{i.e.}
QLA, LQLA, CQLA, and CQLA+) is fixed once an overall mapping has been
determined.  Qalypso, on the other hand, requires the mapper to
calculate how many ancilla generation resources are needed at each
compute and memory region.  The amount of error correction performed
is established in the synthesis phase of our flow
(Figure~\ref{fig:methodology_flow}) and used by the mapper.  
%% However,
%% the precise location of each QEC step is determined by the mapping.
For memory regions, we compute the residency time for each qubit and
use this to automatically add error correction steps when necessary;
the residency calculation permits us to size the amount of ancilla
generation ($M_{ag}$) needed in each memory region.

%% Finally, in \cite{andrew}, the authors show it is not possible to
%% devise an an encoding such that all gates in a universal set may be
%% performed transversally.  

Since it is impossible to perform all encoded gate operations
transversally (\ie~as a simple function of the encoded
bits)~\cite{zeng2007tvu}, we must provide for at least one
non-transversal gate mapping.  Non-transversal gates require special
ancilla generators which are very time- and area-intensive;
consequently, our mapper restricts operation of non-transversal basic
gates to a limited number of locations.

\begin{figure*}
\begin{minipage}{\columnwidth}
    \includegraphics[width=\columnwidth]{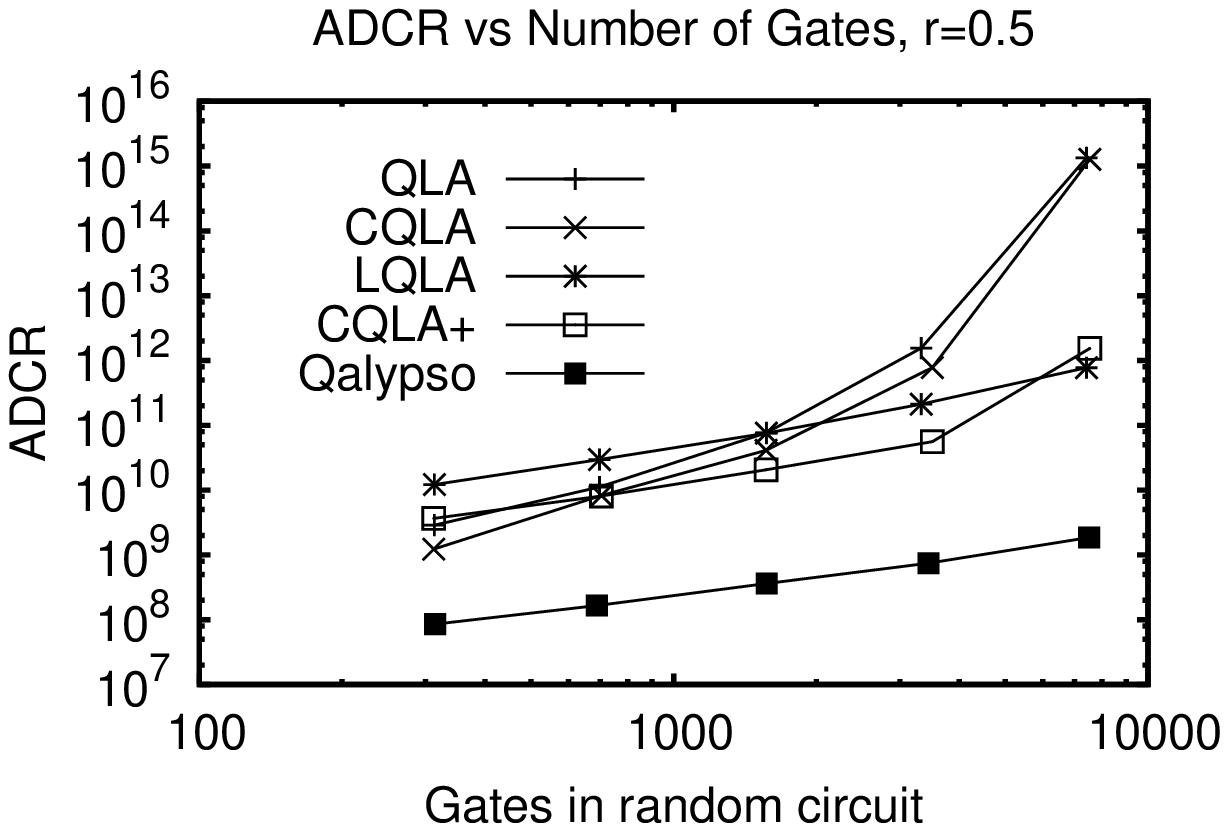}
%,width=\hsize}
\end{minipage}
\begin{minipage}{\columnwidth}
    \includegraphics[width=\columnwidth]{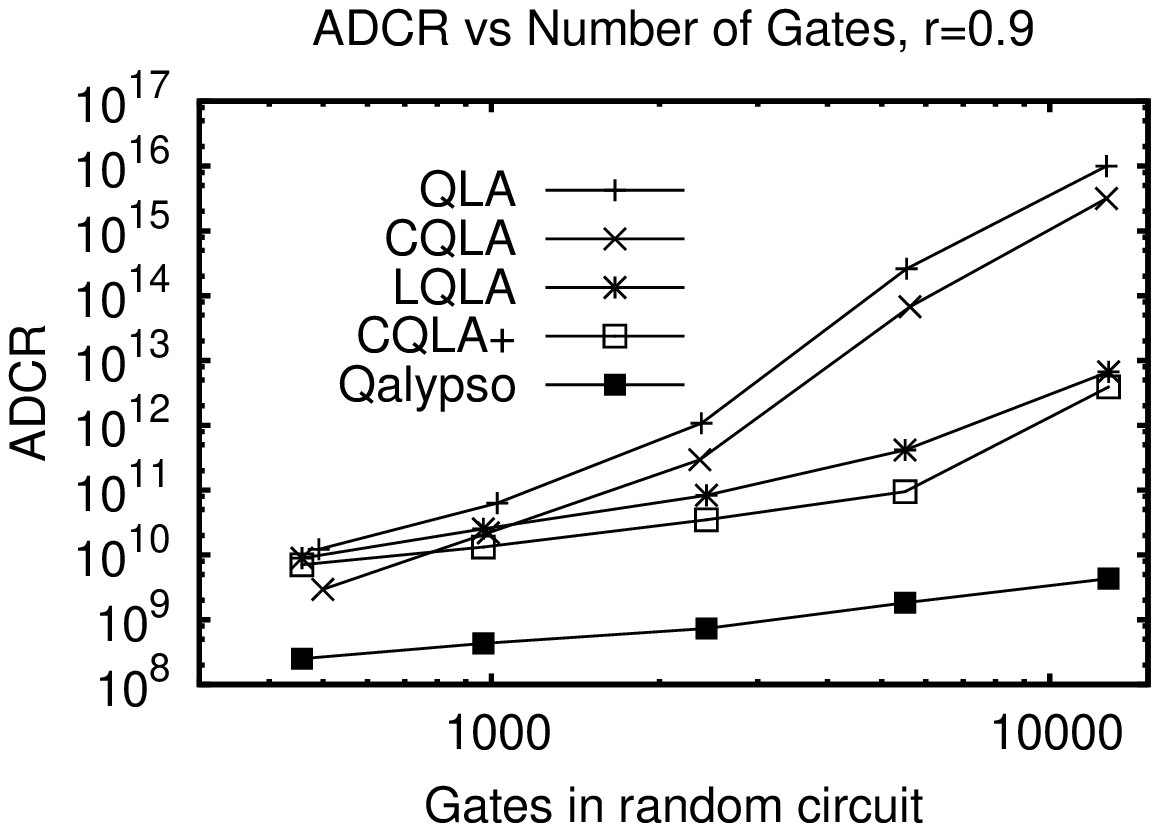}
%,width=\hsize}
\end{minipage}\hfill
\vspace{-0.1in}
\caption{Area-Delay-to-Correct-Result(ADCR) as a function of circuit
  size for the five datapath organizations; the left graph shows
  circuits with Rent's parameter r=0.5 (similar to real circuits), while
  the right shows r=0.9 (more complexity).  We utilize Error Set~2
  (Table~\ref{table:ion_trap_basic_stats}) to emphasize differences.
  Comparison between QLA and CQLA shows the effectiveness of memory
  partitioning mechanisms, while Qalypso shows the advantage of
  the memory partitioning, ancilla balancing, and network resource
  sizing working together.}
\label{plot:random_rent_vs_ADCR}
\end{figure*}

\subsection{Network Resources for Qalypso}\label{sec:networkmap}
As mentioned earlier, teleportation is used to transport data over large
distances (see~\cite{isailovic2006ins, metodi2005qla} for details). The
time to set up a teleportation connection varies according to network
congestion and routing choices.  Ideally, a majority of the setup
latency can be masked by the network, hiding all but the latency of a
single teleportation operation from the data.  In reality, congestion
would limit the ability of dynamic scheduling hardware from realizing
all of this benefit.

During the mapping phase, the mapper assumes all communication can be
done without setup latency.  This assumption is valid only if the
network is sized appropriately.  We size the network by tracking network
consumption during mapping.  Each requested network connection has an
associated path consisting of the source router, intermediate routers,
and destination router. The mapper tracks the load at each router
(number of connections traversing and terminating at the router) for
each time unit of the execution.  When mapping is complete, this load
information is used to determine the area of the router.\footnote{Note
that the \emph{time} component of teleport network usage distinguishes
this channel sizing process from normal wire placement in an ASIC CAD
flow.} The area determined at this phase represents network resources
required to run ``at the speed of the data''~\cite{isailovic2008rqc}.  We
subsequently optimize area usage by retreating from this point; see
Section~\ref{sec:adcr-optimal}.

A high-level floorplan of a network router is shown in
Figure~\ref{fig:router_layout}.  The area of the router consists of
purifiers (P) for EPR purification, teleporters (T) for connections that
span multiple routers, and buffers to store qubits while waiting for the
connection establishment.  The area dedicated to each of these
components is dependent on the maximum load the router sees, as
determined by the mapping phase.  Our tools cannot yet construct a
detailed layout of arbitrarily sized routers.  Instead, in an effort to
obtain realistic network area estimates, we utilize a detailed layout of
a specific sized router to extrapolate the sizes of larger routers.

\subsection{ADCR-Optimal Layouts}\label{sec:adcr-optimal}

Since the inherent parallelism and size of a circuit determines the need
for data and memory regions, some aspect of the mapping process must
select the total area available for mapping.  Further, we need to adjust
the aggressiveness with which the network is sized to meet transient
communication demand.  In principle, we would like to perform the
mapping process with many different configurations, then select designs
which meet some optimization metric.  When the optimization metric is
ADCR (Section~\ref{sec:evaluation}), then we refer to the resulting
circuits as \emph{ADCR-optimal}.  For the remainder of the paper, graphs
which show ADCR should be considered to present ADCR-optimal data
points.

\subsection{Generating Random Circuits to Eliminate Datapath
  Organizations}\label{sec:elimination} 

In this section, we apply our mapping and partitioning techniques,
described above, in order to narrow the candidate datapath organizations
for later sections of the paper.  One strategy for datapath elimination
would be to map real benchmarks to different organizations and use the
result to choose the most effective organizations.  The problem with
this approach is that it requires circuits that vary in size and
complexity to produce a definitive result.  Instead, we map circuits
generated by a random circuit generator that can produce a variety of
sizes and complexities automatically.

The circuit generator produces random gate networks consisting of 1 and
2 qubit gates.  Each random graph is parametrized by a Rent's exponent
\cite{donath1981wld}, which effectively determines the number of wires
crossing recursive min-cut partitions of the gate network; a larger
Rent's exponent signifies a circuit with greater connectivity.  While
other works have discussed random gate network generation for classical
circuits \cite{darnauer1996mgr}, the specific fan-in and
fan-out requirements for quantum circuits preclude these solutions.
Instead, we pick the number of gates and qubits desired, then, for each
gate, randomly pick 1 or 2 qubits and place the gate at the end of the
network.  We generate many of these random networks and plot the results
as a function of gate count.

Figure~\ref{plot:random_rent_vs_ADCR} shows the cumulative effects of
our partitioning mechanisms, ancilla balancing and network resource
sizing by plotting the composite ADCR metric (see
Section~\ref{sec:evaluation}) using Error Set 2
(Table~\ref{table:ion_trap_basic_stats}) as a function of circuit size
for random circuits.  For larger circuits, we see that the QLA and CQLA
datapath organizations are far less area-efficient than the others.
Consequently, the remainder of the paper will focus on mapping to LQLA,
CQLA+, and Qalypso.
% LocalWords:  Datapaths

\begin{figure*}[t!]
 \begin{center}
    \includegraphics[width=\textwidth]{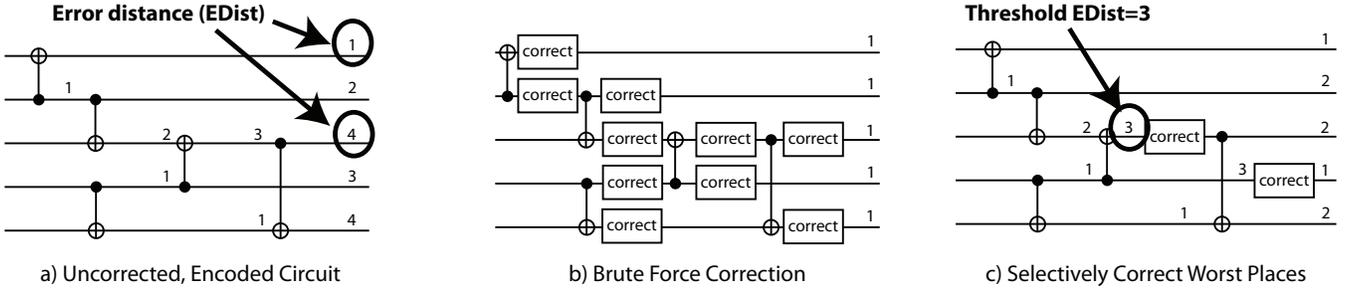}
%,width=0.95\hsize}
  \end{center}
  \vspace*{-0.2in}
  \caption{Error Correction as Retiming: (a) Simple model of counting
    errors: each gate adds one error unit to each qubit involved, while
    interacting qubits propagate error counts to each other.  (b)
    Standard conservative placement of error correction: one
    after every gate, while (c) Retiming Optimization:  Place fewer
    corrections while maintaining similar qubit error counts to
    dramatically reduce total gate count.}\label{fig:correction_opt}
%  \vspace*{-0.05in}
\end{figure*}

\vspace{0.1in}
\section{Optimizing QEC}\label{sec:qec}

In this section, we tackle the misconception that quantum error
correction should occur after every quantum gate
\cite{preskill1997ftq}. In fact, the insertion of quantum error
correction operations can actually serve to \emph{increase} the
probability of error if used indiscriminately. The reason for this
phenomenon is that quantum error correction involves gate operations
and data movement -- each of which increase the probability of error
-- even as they work to suppress this error.

The goal of the QEC synthesis flow discussed in this section is to
perform \emph{selective} error correction, as originally suggested in
\cite{oskin2002par}.  This paper provides the first concrete
method for selectively inserting error correction operations into an
encoded quantum circuit.
%% Our optimized error
%% correction insertion algorithm differs from all previous proposals
%% which suggest adding error correction operations after every gate in
%% the circuit. 
Through mapping, partitioning, layout, and error simulation, we evaluate
the resulting circuits to show that we can vastly improve area and
latency while improving error behavior.

\subsection{Casting Error Correction as Retiming}

The standard ``brute force'' error correction method, which adds an
error correction step after every operation, is unnecessary in many
circuits because not every qubit operation contributes the same amount
of error:  
%% The reason is that qubits are affected by errors in other
%% qubits it interacts with.  
a qubit that interacts with many ``dirty'' qubits will accumulate errors
much faster than a qubit that interacts with ``clean'' qubits.  The crux
of our technique is to estimate the \emph{critical error path}, or the
sequence of qubit interactions that introduces the most error into the
circuit, in a fashion that is similar to the critical latency path
through a circuit.  We will focus error correction resources on these
critical error paths.

Figure~\ref{fig:correction_opt}a illustrates the process of estimating
the critical error path.  It uses a simple, but effective model of a
complicated error propagation process to estimate a parameter we refer
to as the \emph{Error Distance} (EDist).  This model assumes that (1)
each gate introduces one unit of error and (2) all qubits interacting
within a gate acquire the maximum error value out of those qubits.  In
this simple model, the correction procedure resets the error counts of
the corrected qubits to a base value (the correction process itself is
noisy so it is not typically set to 0).  We assume an error correction
can only work if an incoming qubit's error count is below some
threshold.
%%  (in Figure \ref{fig:correction_opt}a, the threshold is 3).

The traditional brute force method, shown in
Figure~\ref{fig:correction_opt}b, conservatively applies error
correction after every gate to ensure that qubit errors do not propagate
to other qubits.  In our optimized process, shown in
Figure~\ref{fig:correction_opt}c, we only add error correction steps
after operations whose EDist value exceed a critical threshold (in
the figure, the threshold is 3).

We have formulated this optimization problem as a case of circuit
retiming \cite{leiserson1983osc} from the classical CAD literature.  We
choose where to insert correction steps by solving the
Bellman-Ford-style circuit retiming problem described in
\cite{leiserson1983osc}.  More specifically, we use the \emph{OPT2}
clock period minimization from \cite{leiserson1983osc} by replacing the
clock period with the maximum, tolerable error count (expressed as a
threshold value of EDist).  We must make one modification to
the algorithm: our initial network has no ``registers'' (error
correction steps), so we must perform a binary search to determine the
number of correction steps are needed to achieve a particular
period/error count.

Additionally, there is no constraint on the
distribution of correction steps as there is on synchronous registers in
the retiming case.  This give us much more freedom to distribute them.
In fact, a valid optimized placement might mean a particular qubit does
not get corrected at all, if it only goes through a few gates.

\begin{table}
  \begin{tabular}{|l|r|l|r|}
    \hline
    Correction? & EDist & Output success & Operation count\\
    \hline\hline
    Every gate & N/A & 0.987 & 3105611\\
    \hline
    Optimized & 3 & 0.966 & 708811\\
    \hline
    Optimized & 6 & 0.956 & 366411\\
    \hline
    Optimized & 9 & 0.870 & 24011\\
    \hline
    None & N/A & 0.65 & 1024\\
    \hline
  \end{tabular}
  \vspace*{-0.1in}
  \caption{Performance of QEC optimization as function of EDist.  Output
    success is probability that output data is not corrupted by an
    uncorrectable error, using only gate errors.  Operation count is the
    total number of physical gates in  entire
    circuit. }\label{table:error_and_counts}
  \vspace*{-0.1in}
\end{table}

Table~\ref{table:error_and_counts} illustrates why this error
optimization seems so promising at first glance.  It shows the results
of optimizing an \emph{unmapped}, 1000-bit random circuit using a variety of
EDist values.  We see that an optimized circuit can demonstrate little
or no degradation in the probability of success for factors of 2 or 3 in
operation reduction.  Figure ~\ref{plot:random_graphs_edist_vs_mapped}
shows that the probability of a \emph{mapped} circuit actually improves under
some circumstances with the optimization -- even though the optimization
is performed on the unmapped circuit.

By tuning the EDist threshold, we adjust the aggressiveness of the
optimization.  To do this automatically, we can perform a binary search
over possible thresholds and pick (1)~the count with the highest success
probability, (2) the highest value that achieves a minimum success
probability, or (3) the best success probability within a particular
resource budget.  While our simple EDist propagation model is useful
while performing the optimization, it is important to evaluate the
resulting circuits with a more accurate model --- after mapping.  Thus,
one should iterate through the layout process as part of choosing an
optimal EDist value.

\subsection{Validation of Retiming Heuristic}

 \begin{figure}
 \vspace*{-0.1in}
\begin{center}
 \includegraphics[width=\columnwidth]{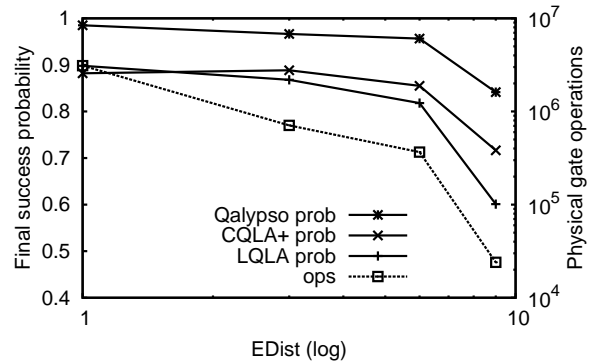}
%height=2in}
\end{center}
 \vspace{-0.25in}
 \caption{Post-Mapping Success Probability and Physical Operation Count
   as a function of EDist for a 3500 gate random graph (r=0.5).
%%Note
%%   that EDist only impacts the number of correction operations in the
%%   logical circuit and therefore is independent of architecture.
% Error is normal for this figure, 6,8,10
%The error here include the low move ($1*10^{-10}$) and idle error ($1*10^{-12}$).
}
 \vspace*{-0.1in}
 \label{plot:random_graphs_edist_vs_mapped}
 \end{figure}

 \begin{figure}
 \vspace*{-0.05in}
 \includegraphics[width=\columnwidth]{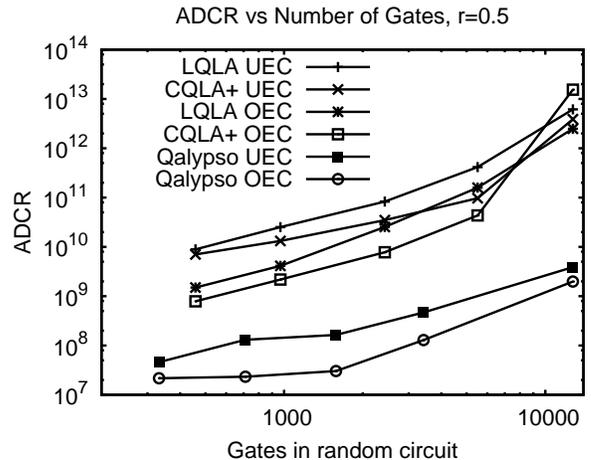}
%,height=2.5in}
 \vspace{-0.15in}
 \caption{ADCR for optimized and unoptimized random graphs (r=0.5),
 mapped to the three most promising datapath organizations
 (Section~\ref{sec:elimination}), using Error Set~2
 (Table~\ref{table:ion_trap_basic_stats}) for added difficulty.  
% Error is normal for this figure, 6,8,10
%The error here include the low move ($1*10^{-10}$) and idle error ($1*10^{-12}$).
}
 \vspace*{-0.1in}
 \label{plot:optimized_random_graphs_gates_vs_ADCR_5}
 \end{figure}

For the remainder of the paper, we choose option (2), from above:
perform a binary search to determine the value of the EDist parameter
that yields no more than a 5\% degradation in the probability of success
on a mapped circuit (\eg, optimization along the X-axis of
Figure~\ref{plot:random_graphs_edist_vs_mapped}).

Figure~\ref{plot:optimized_random_graphs_gates_vs_ADCR_5} presents the
results of optimizing random circuits with Rent's parameter r=0.5 on
the three most promising datapath organizations from
Section~\ref{sec:elimination}, using Error Set~2
(Table~\ref{table:ion_trap_basic_stats}) for emphasis.  This figure
shows that the QEC optimization is quite successful in reducing ADCR on
mapped circuits, showing up to a factor of 10 improvement in many cases.
Section 5 shows that the optimization works even better on non-random
circuits, such as adders.

%In subsequent sections, we show
%that this optimization works extremely well for non-random topologies
%such as adders and factoring circuits.

\subsection{Post-Mapping Adjustments}

Our QEC optimization utilizes a simplistic gate error model during the
placement of error correction operations.  Since it effectively works on
\emph{unmapped} circuits, it does not currently adjust for other sources
of error.  Although the previous section showed that this technique
performs quite well on \emph{mapped} circuits, we could imagine that
high movement and idle error rates or long channels in a large circuit
could lead to a mapped circuit that behaves sub-optimally after
optimization.  Specifically, we expect to see two effects:
\begin{itemize}
\item Degradation in fault tolerance as the gate error based
  optimization becomes less relevant due to relative increases in
  other error types.
\item Improvement in fault tolerance as reduction in error correction
  resources leads to more compact designs with shorter communication
  distances.
\end{itemize}
%% These two conflicting scenarios make it difficult to predict how error
%% correction optimization will perform.  
As future work, we anticipate reevaluating the error propagation of a
circuit, once it has been mapped, and adjusting the mapped circuit
through selective addition or removal of error correction operations.
This \emph{post-mapping QEC adjustment} will remove non-uniform error
propagation introduced by the mapping process.  Hopefully this step
would lead to a higher probability of success.

\begin{figure*}[t!]
\vspace*{-0.1in} 
\begin{minipage}[b]{\columnwidth}
\begin{center}
\vspace*{-0.2in} 
\includegraphics[width=\columnwidth]{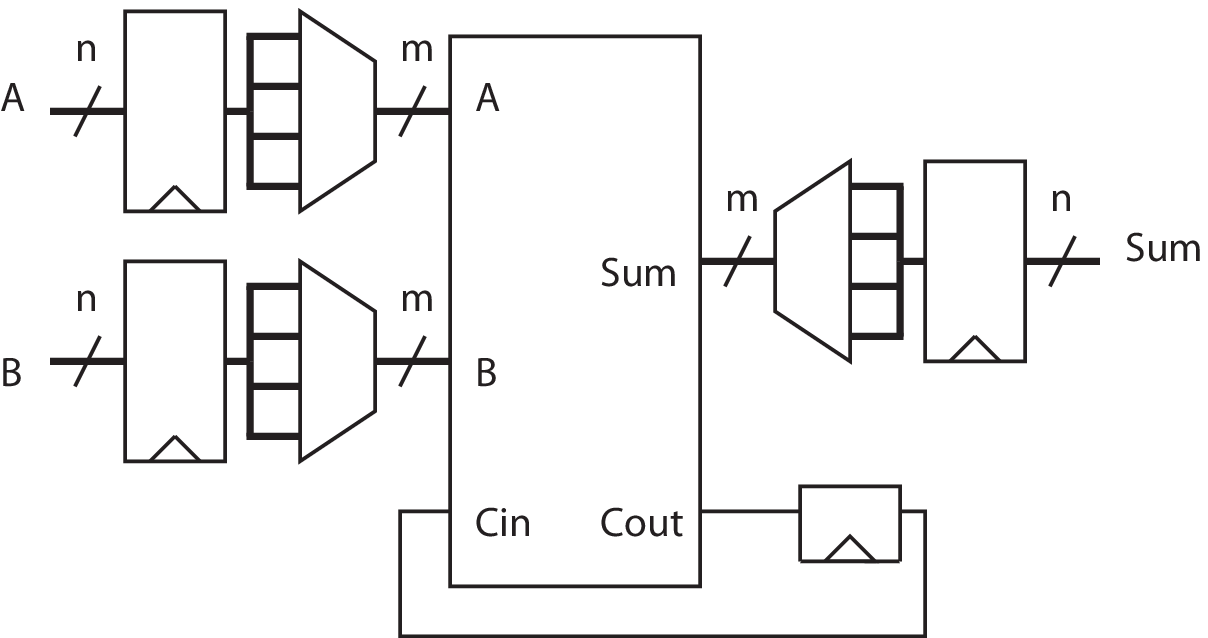}
%,width=0.95\hsize}
\vspace*{-0.15in}
\end{center}
\end{minipage}\hfill
\begin{minipage}[b]{\columnwidth}
\begin{center}
\includegraphics[width=\columnwidth]{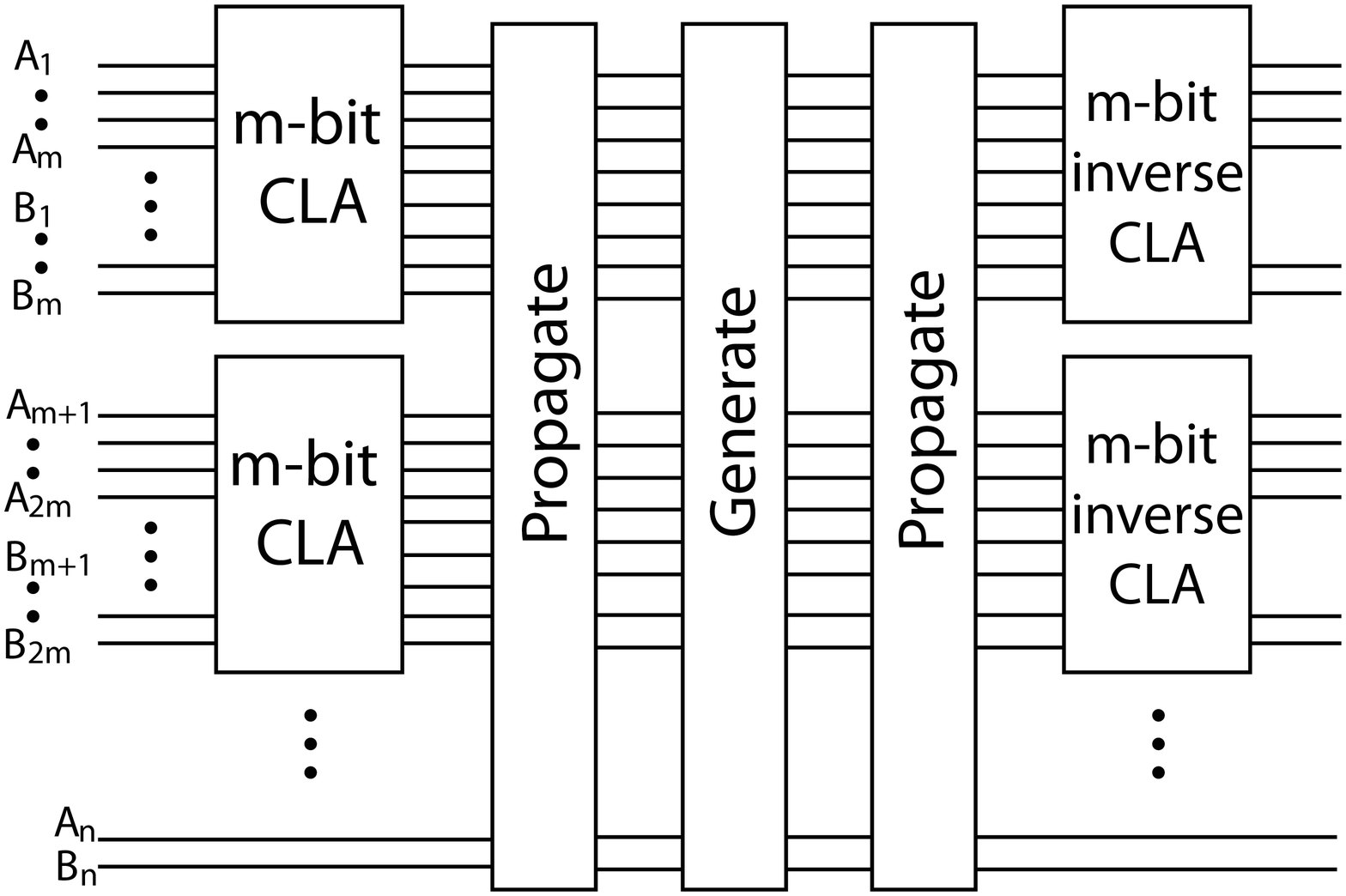}
%,width=0.875\hsize}
\vspace*{-0.25in}
\end{center}
\end{minipage}
\begin{minipage}[t]{\columnwidth}
\caption{Our Quantum Ripple-Carry Adder (QRCA):~An $n$-bit ripple-carry
 adder using $\frac{n}{m}$ cycles through an $m$-bit ripple carry
  adder.  
%Carry out is registered and cycled back into subsequent
%  iterations.  
Each ripple-carry block is similar to a classical ripple
  carry except that the carry bit must be inverted at the end to
  disentangle ancilla qubits.}
\vspace*{-0.05in}
\label{fig:ripple_carry}
\end{minipage}\hfill
\begin{minipage}[t]{\columnwidth}
\caption{Our Quantum Carry-Lookahead Adder (QCLA):~As with a
  classical CLA, the first few levels of the propagate and generate
  networks are built with uniform sized blocks.  
%The logarithmic
%  depth networks are completed in the blocks that span all bits.  
We must reverse the propagate and generate bits to disentangle them from
  the output.}
\vspace*{-0.05in}
\label{fig:cla}
\end{minipage}
\begin{minipage}[b]{\columnwidth}
\begin{center}
 \includegraphics[width=\columnwidth]{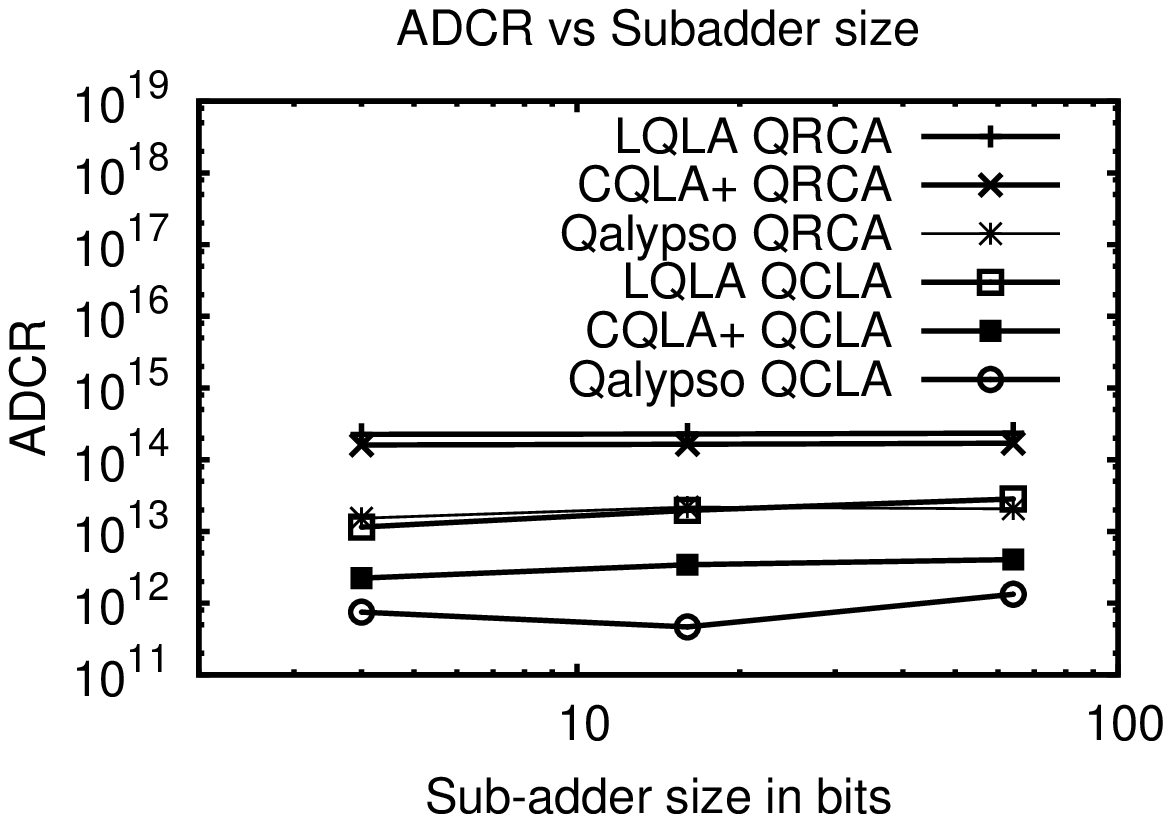}
%,width=\hsize}
\vspace*{-0.25in}
\end{center}
\end{minipage}\hfill
\begin{minipage}[b]{\columnwidth}
\begin{center}
 \includegraphics[width=\columnwidth]{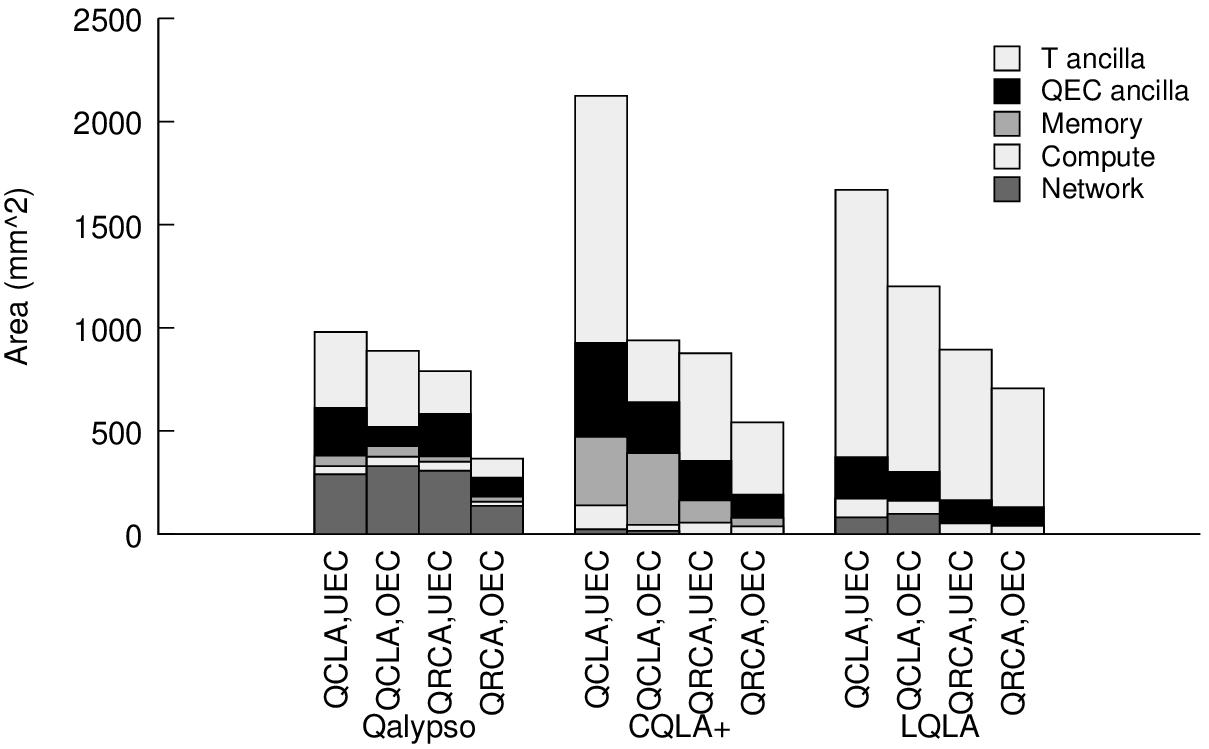}
%,width=\hsize}
\vspace*{-0.15in}
\end{center}
\end{minipage}
\begin{minipage}[t]{\columnwidth}
\vspace*{-0.1in}
 \caption{ADCR for 1024-bit QRCA and QCLA adders \emph{with QEC
   optimization} (Section~\ref{sec:qec}) as the sub-adder 
   size is varied.  Using Error Set~1 (Table~\ref{table:ion_trap_basic_stats})}
\vspace*{-0.1in}
 \label{fig:optimized_cla_vs_ripple_ALCR}
\end{minipage}\hfill
\begin{minipage}[t]{\columnwidth}
\vspace*{-0.1in}
 \caption{Area breakdowns for both unoptimized (UEC) and optimized
   (OEC) versions of 1024-bit adders with three datapath organizations.}
\vspace*{-0.1in}
 \label{plot:adder_area_breakdown}
\end{minipage}
\end{figure*}

\section{Adder Microarchitectures}\label{sec:adderdesign}

As discussed in Section~\ref{sec:intro}, the quantum adder is a
fundamental component of Shor's factorization.  Consequently, this
section will apply the machinery that we developed in previous sections
to produce optimized adder circuits.  Since we are targeting 1024-bit
factorization, we will examine 1024-bit adders.  Further, for non-random
circuits, we will switch to the more realistic Error Set~1 from
Table~\ref{table:ion_trap_basic_stats}.

%% This functional unit is just as necessary to a quantum
%% machine as to a classical machine.
%% For this reason, we use this key functional unit as our benchmark for
%% evaluating our error correction optimization and our mapping
%% heuristics.

\subsection{Ripple-Carry vs Carry-Lookahead}\label{sec:adderchoice}
We evaluate the quantum ripple-carry adder (QRCA)~\cite{draper2000aqc}
and the quantum carry look-ahead adder (QCLA)~\cite{draper2004ldq},
constructing larger adders from smaller adder modules, similar to what
is done with classical bit-serial adders.
%% , although we can have more than one instance of the
%% smaller adder in the datapath.  
Figure~\ref{fig:ripple_carry} shows how
an $n$-bit QRCA is constructed with multiple passes through a single
$m$-bit sub-adder. The registers can map to memory regions and the adder
block to compute regions such that data is shuttled between memory and
compute when it uses the sub-adder.  Similarly, Figure~\ref{fig:cla}
shows how an $n$-bit QCLA is constructed with smaller modules.  The
modular approach allows us to trade area for parallelism thus allowing us
to construct optimal adder configurations.

Figure~\ref{fig:optimized_cla_vs_ripple_ALCR} shows ADCR as a function
of sub-adder size for both QRCA and QCLA architectures.  This figure
shows QEC-optimized circuits mapped to the three best datapath
architectures.  Interestingly, QCLA beats QRCA by a factor of 10 or 20
(for a given datapath organization).  Since QCLA seems to be so much
more area-efficient, we will concentrate the bulk of our analysis on
QCLA.

\begin{figure*}[t!]
\begin{minipage}[b]{\columnwidth}
\begin{center}
\vspace*{-0.1in}
\includegraphics[width=\columnwidth]{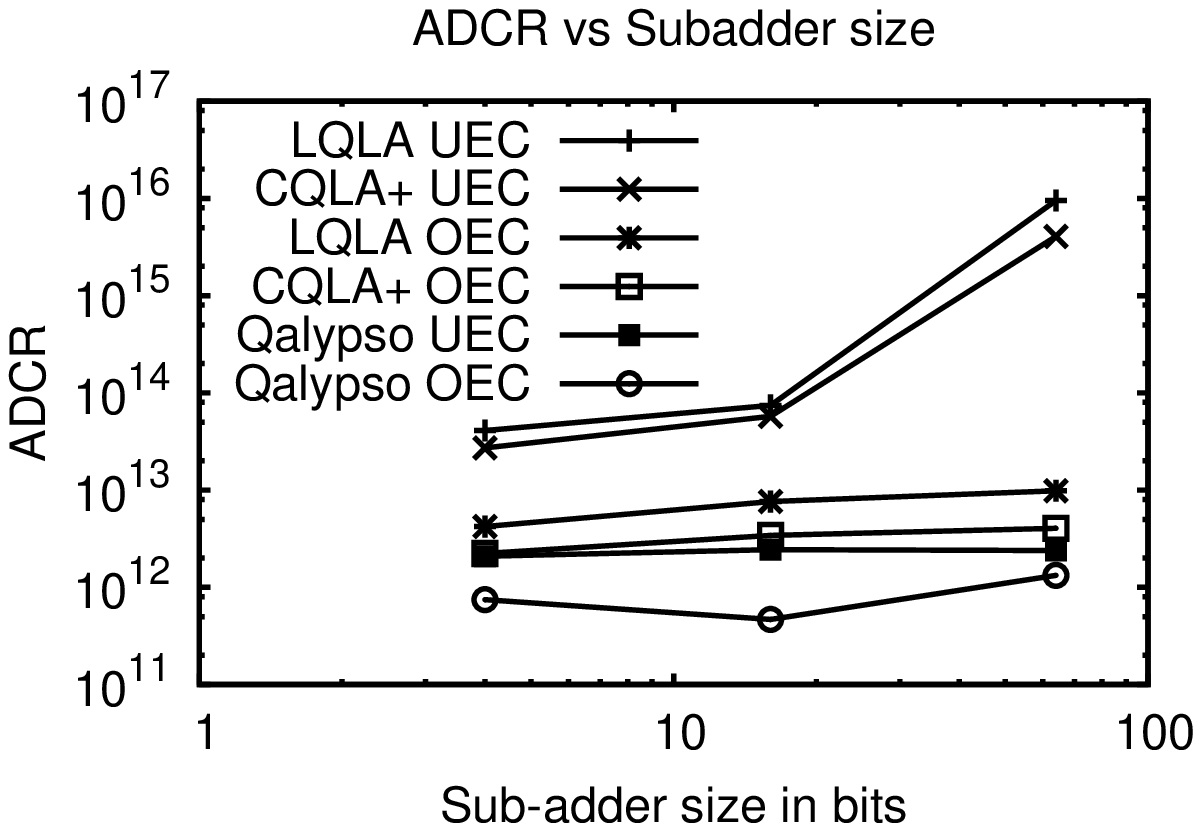}
%,width=\hsize}
\vspace*{-0.4in}
\end{center}
\end{minipage}\hfill
\begin{minipage}[b]{\columnwidth}
\begin{center}
\vspace*{-0.1in}
\includegraphics[width=\columnwidth]{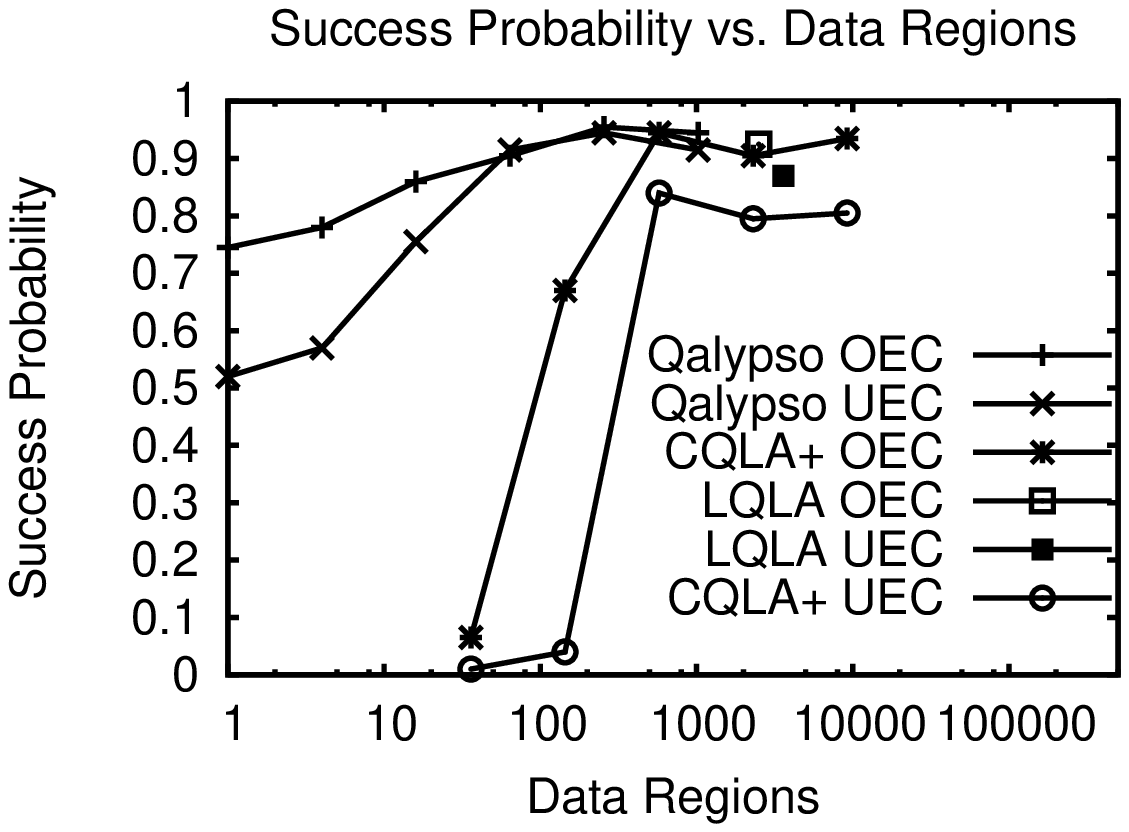}
%,width=\hsize}
\vspace*{-0.4in}
\end{center}
\end{minipage}
\begin{minipage}[t]{\columnwidth}
 \caption{Comparison of optimized (OEC) and unoptimized (UEC) QEC on 1024-bit QCLA as a
 function of sub-module size, using Error Set~1 (Table~\ref{table:ion_trap_basic_stats}).   }
 \label{fig:effectiveness_qec_optimized_cla_1024}
\vspace*{-0.1in}
\end{minipage}\hfill
\begin{minipage}[t]{\columnwidth}
 \caption{Success probability for 1024-bit QCLA as we vary the
   number of data regions, using Error Set~1 (Table~\ref{table:ion_trap_basic_stats}).}
 \label{fig:optimized_cla_fus_vs_prob}
\vspace*{-0.1in}
\end{minipage}
\end{figure*}

\subsection{Area for Quantum Error Correction}

One of the more interesting issues that we can tackle with our CAD
flow is to determine what fraction of an optimized circuit is devoted
to quantum error correction.  Some authors have contended that error
correction is \emph{the} dominant activity of a quantum computing
circuit~\cite{steane2004bbg,eval_framework,thaker2006qmh,metodi2005qla}.

Figure~\ref{plot:adder_area_breakdown} belies this conclusion.  It shows
an area breakdown for ADCR-optimal circuits for both QRCA and QCLA, with
and without the QEC optimization from Section~\ref{sec:qec}.  The only
area truly devoted to error correction is the ``QEC ancilla'' (producing
zero-ancilla for use in error correction).  The memory area (``Memory'')
accounts for the storage of bits, but not the zero-ancilla generation
required for error suppression (which is included in the QEC ancilla
category).  Two crucial components often ignored by other authors are
the area devoted to the teleportation network (``Network'') and the area
devoted to T-ancilla (``T ancilla'').  This latter category is required
to perform non-transversal quantum gates and is present in many real
circuits\footnote{Although one might argue that some fraction of the
T-ancilla area is performing error correction as part of generating
T-ancilla for LQLA and CQLA+, T-ancilla generation does not strictly
\emph{require} error correction.}.

We can glean several interesting results from
Figure~\ref{plot:adder_area_breakdown}: First, the area devoted to error
correction (QEC-ancilla) is only 20\%-40\% for optimized Qalypso designs
(where ancilla generation can be effectively balanced).  Second, the QEC
optimization reduces the QEC-ancilla generation by almost half for
Qalypso and the combined QEC-ancilla and T-ancilla generation by more
than half for other datapaths.

It is interesting to compare area utilization with operation counts: for
the QEC-optimized, Qalypso mapped, 1024-bit QCLA, we find that 70\% of
the total operations are devoted to QEC ancilla generation, but only 5\%
are devoted to error correction (interaction of ancilla with data).
Thus, the pipelined ancilla factories are working at high utilization to
feed the actual error correction operations, suggesting that raw
operation count is not a particularly good metric for understanding the
overhead of error correction.

\subsection{Studying the QCLA Architecture}\label{sec:study_qcla}

Section~\ref{sec:adderchoice} showed that the QCLA architecture is the
clear winner over QRCA in area-efficiency (ADCR).  We will
investigate this architecture a bit more in this section.

First, Figure~\ref{fig:effectiveness_qec_optimized_cla_1024} shows the
effectiveness of the error correction optimization on the quantum carry
look-ahead adders.  We see that the QEC-optimization can be extremely
effective---lowering ADCR by a factor of 46 for the optimal Qalypso
design and 2 to 3 orders of magnitude for the other datapath
organizations.  These latter datapath organizations suffer from
serialization on ancilla generation and see a great improvement in
latency as a result of optimization.

 \begin{figure}
 \vspace*{-0.25in}
 \begin{center}
 \includegraphics[width=\columnwidth]{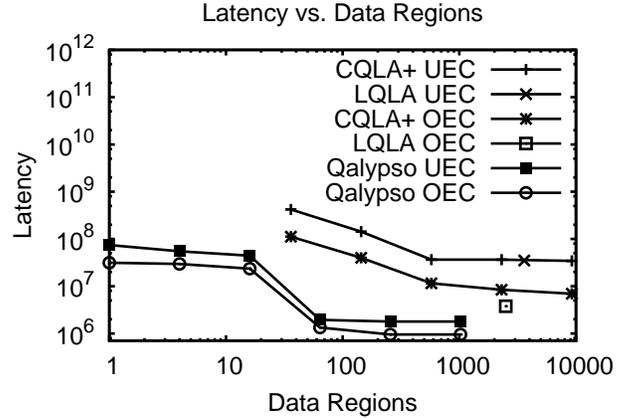}
%,width=\hsize}
 \end{center}
 \vspace{-0.3in}
 \caption{Latency for 1024-bit QCLA as we vary the number of data
   regions. }
 \vspace*{-0.1in}
 \label{fig:optimized_cla_fus_vs_latency}
 \end{figure}

Second, Section~\ref{sec:adcr-optimal} discussed the need for varying
the configurations considered by the mapper in order to find an optimal
layout.  Figures~\ref{fig:optimized_cla_fus_vs_prob}
and~\ref{fig:optimized_cla_fus_vs_latency} show how important it is to
perform parameter variation during layout: they show the sensitivity of
the QCLA layout to the number of data regions given to the mapper.  As
can be seen by these figures, both success probability and latency are
strongly impacted when the mapper is given too few data regions.  Beyond
a certain point however, the mapper is able to achieve high probability
of success and low latency.  The ADCR-optimization seeks to find the
knee of the curve, namely the lowest area that achieves low latency and
high probability of success.

\begin{figure*}[t!]
  \begin{center}
    \includegraphics[width=0.95\textwidth]{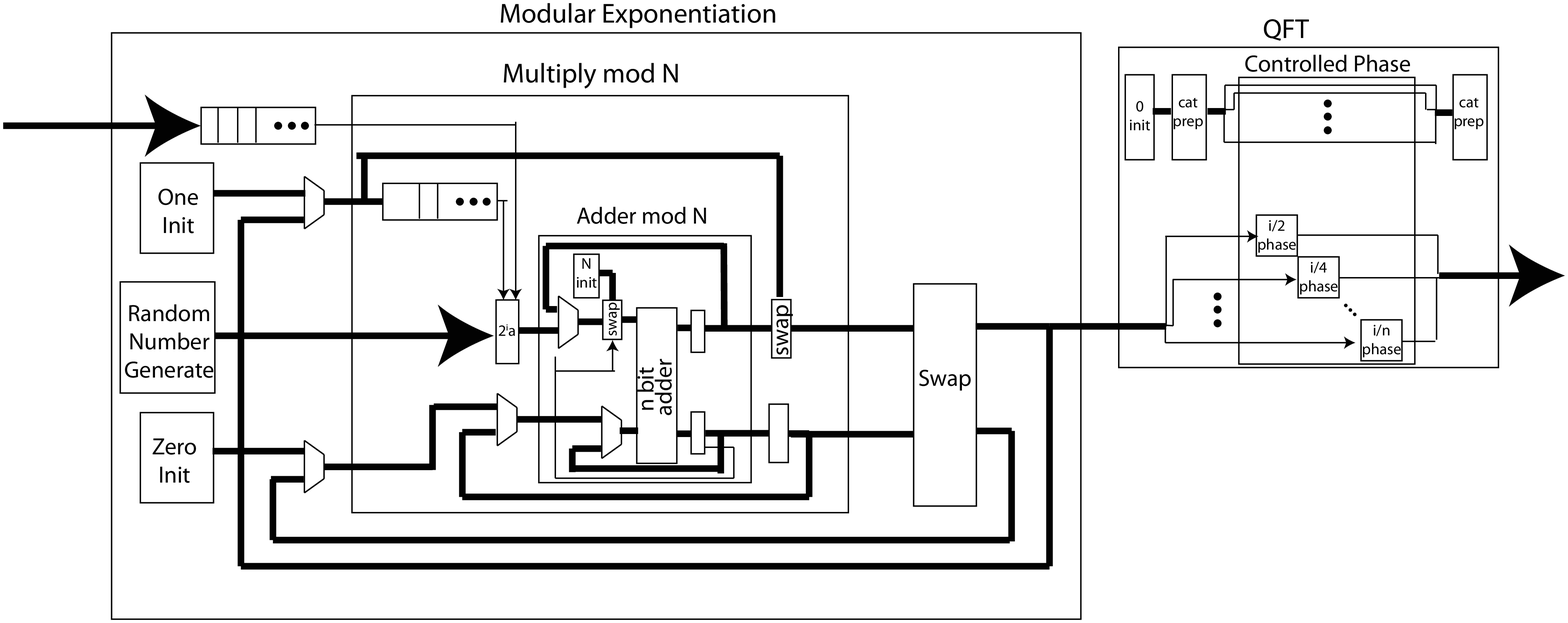}
  \end{center}
  \vspace*{-0.25in}
  \caption{Shor's factorization consists of two major phases: modular
    exponentiation and Quantum Fourier Transform.  The modular
    exponentiation circuit comprises the bulk of the execution time for
    Shor's factoring.}
  \label{fig:ShorComplete}
  \begin{minipage}[b]{\columnwidth}
  \begin{center}
    \includegraphics[width=\columnwidth]{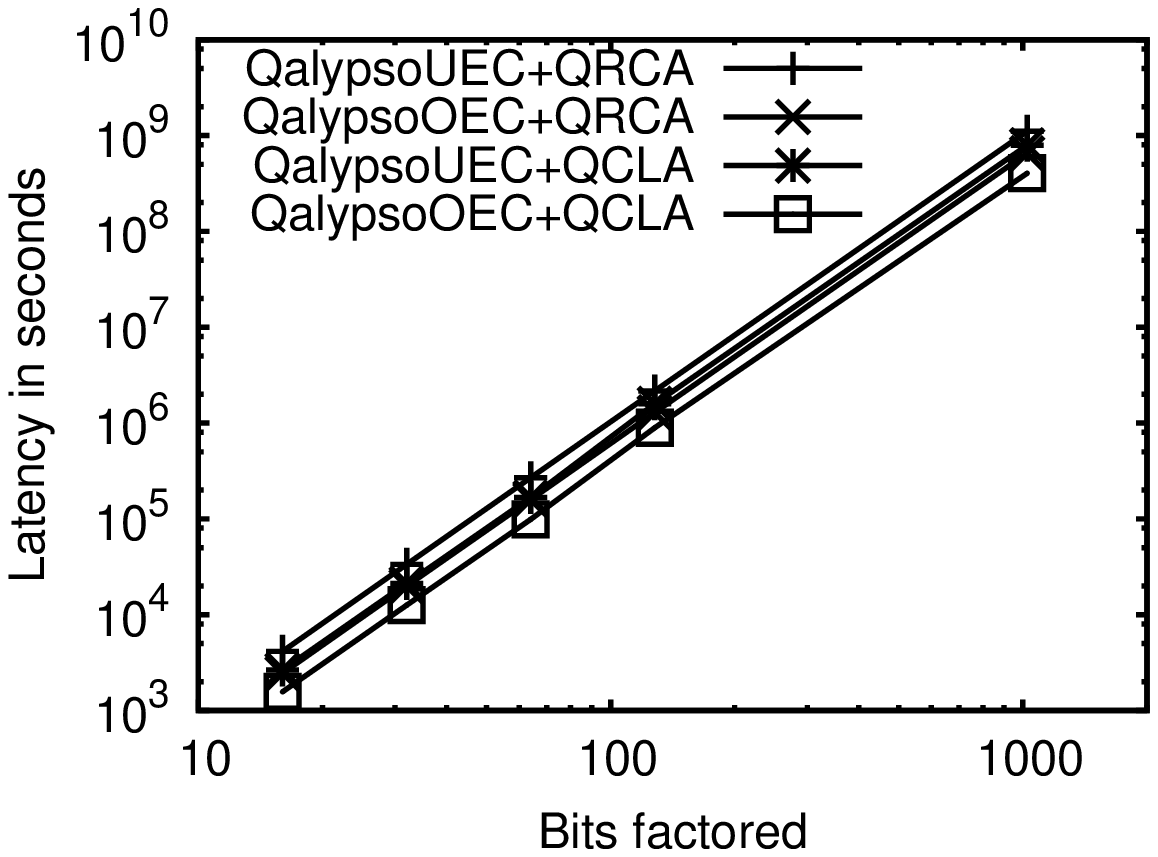}
%,width=\hsize}
  \end{center}
  \end{minipage}\hfill
  \begin{minipage}[b]{\columnwidth}
  \begin{center}
    \includegraphics[width=\columnwidth]{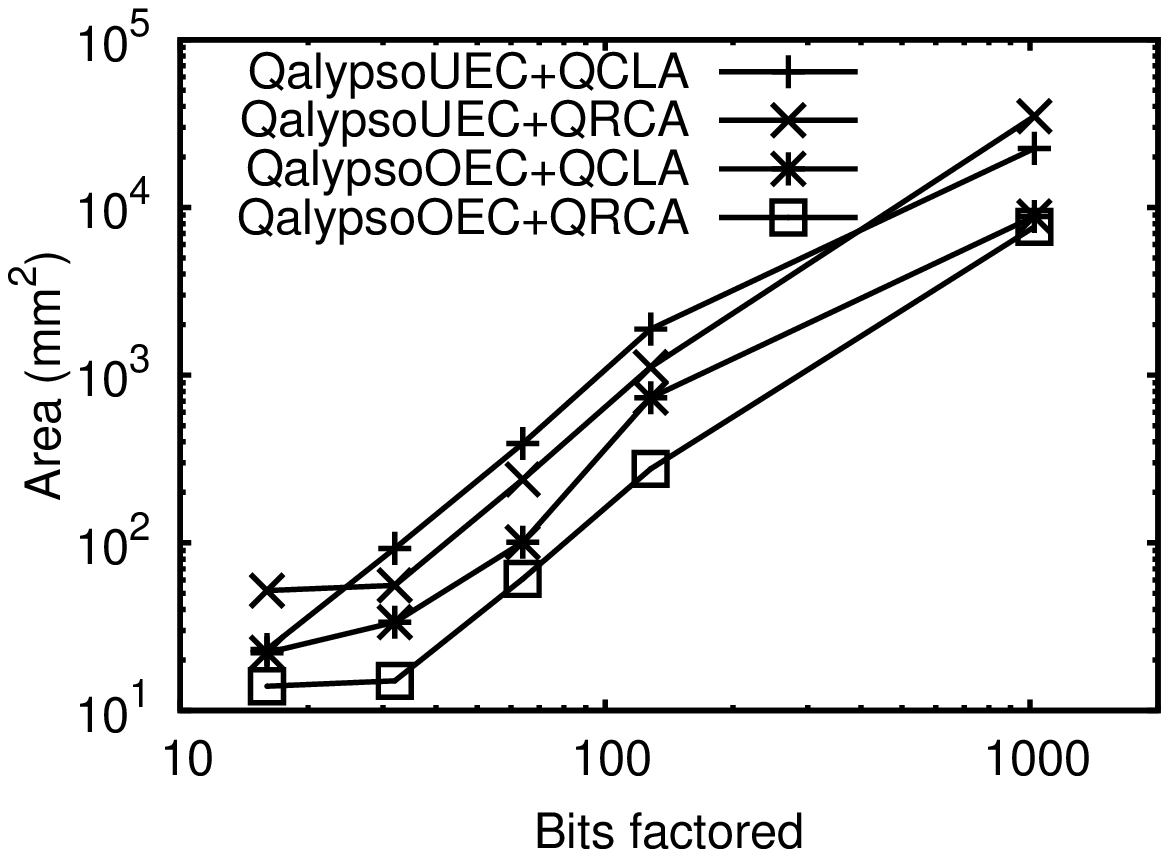}
%,width=\hsize}
  \end{center}
  \end{minipage}
  \begin{minipage}[t]{\columnwidth}
  \vspace*{-0.2in}
  \caption{Total runtime of Shor's factoring algorithm as a function
    of the number of bits factored.}
  \label{fig:shor_bits_vs_latency}
  \vspace*{-0.05in}
  \end{minipage}\hfill
  \begin{minipage}[t]{\columnwidth}
  \vspace*{-0.2in}
  \caption{Total area used by Shor's factoring algorithm as a function
    of the number of bits factored.}
  \vspace*{-0.05in}
  \label{fig:shor_bits_vs_area}
  \end{minipage}
\end{figure*}

\section{1024-Bit Shor's Factorization}\label{sec:shor}

%% \begin{table}
%%   \begin{tabular}{|c|c|c|c|}
%%     \hline
%%     & Area & Latency & Success Rate \\
%%     \hline
%%     QADD & 6304 & 40828604 & 0.1945 \\
%%     CQLA & 26874 & 9786950 & 4.05e-161 \\
%%     \hline
%%   \end{tabular}
%%   \caption{Shor's algorithm on a 4 bit number}
%% \end{table}

%% \begin{figure}
%% \begin{minipage}{0.48\hsize}
%% \begin{center}
%% \epsfig{file=graphs/shor_latency.eps,width=\hsize}
%% \end{center}
%% \vspace{-0.2in}
%% \caption{Total latency for performing Shor's factorization algorithm
%% for varying sized numbers.}
%% \label{fig:shor_latency}
%% \end{minipage}
%% \hfill
%% \begin{minipage}{0.48\hsize}
%% \begin{center}
%% \epsfig{file=graphs/shor_area.eps,width=\hsize}
%% \end{center}
%% \vspace{-0.2in}
%% \caption{Total area of a datapath for performing Shor's factorization
%% algorithm for varying sized numbers.}
%% \label{fig:shor_area}
%% \end{minipage}
%% \end{figure}

With Section~\ref{sec:adderdesign} as a prelude, we are now ready to
compute the latency and area of optimal Shor's factoring
circuits\footnote{Our failure probability simulation is not yet up to
handling circuits of the size of Shor's factoring; this is future
work.}.  We will use the Qalypso datapath organization, since it has
shown consistent advantage for both random circuits \emph{and} adder
circuits.  Figure \ref{fig:ShorComplete} shows a block-diagram of our
target circuit.  It consists of two main components: modular
exponentiation and the quantum Fourier transform (QFT).  For the modular
exponentiation circuit, we rely on the work done in \cite{vedral1996qne}
and for the QFT, \cite{fowler2004sss}. Since addition is a key component
of the modular exponentiation circuit, we use our best adder designs
from Section~\ref{sec:adderdesign}.
%% in terms of sub-adder size and data
%%region count.

%% Similar to the
%% adders shown in the previous section, we synthesize designs to factor
%% numbers up to 1024 bits in size using Shor's algorithm, shown in Figure
%% \ref{fig:ShorComplete}.  

Our full, unoptimized implementation of Shor's algorithm, including
error correction, consists of $1.35 \times 10^{15}$ physical operations.
Our optimized version has \emph{1000x fewer} physical operations.  The
majority of the computation is dedicated to modular exponentiation
(computing $a^x\mod N$); much less than 1\% of the computation is spent
on the QFT.

In Figure~\ref{fig:shor_bits_vs_latency}, we see that the QEC
optimization yields a 55\% improvement in latency for all circuit
sizes using QCLA.  Further, the difference in latency between
factoring with QCLA and QRCA is about 40\%.  The fastest time for
factoring 1024 bits is $6\times 10^8$ seconds, using a QCLA version of
Shor's.  We note that this runtime is quite long; in the future, we
hope to reduce this time with a more parallel implementation or
further circuit optimizations.

%% We consider it future work to try
%% to reduce this by using a more parallel factoring implementation and
%% further circuit optimizations to expose parallelism.

Figure~\ref{fig:shor_bits_vs_area} shows area as a function of bits
factored.  One noticeable result is that the smallest area for 1024-bit
factoring is 7659~mm$^{2}$ using QRCA; this is more than 2 orders of
magnitude smaller than previous estimates~\cite{metodi2005qla}.  Another
result is that the QEC optimization gives a 2.6x improvement in area for
the QCLA version.  The optimization nearly closes the area gap between
QRCA and QCLA versions at 1024 bits (only a 13\% difference).  There are
two possible causes for this: (1) we have not searched enough of the
configuration space to pick the best QRCA design or (2) the primary
error path in the QRCA (the long carry chain) offers only limited
opportunities for optimization.
%%The QCLA has a much more
%%complex structure and thus the number of opportunities increases with
%%adder size.

If we look at the breakdown in area, we see that the 1024-bit
unoptimized QCLA-based design has 21\% of its area dedicated to
QEC-ancilla whereas the optimized version uses 12\% of its area for QEC.
Further, the QRCA-based design has about $\frac{1}{2}$ the QEC-ancilla
area of the QCLA-based one.
%although the area reduction due to optimization is less dramatic.

%% First, the success rate of the algorithms graphed in
%% Figures~\ref{fig:shor_latency} and~\ref{fig:shor_area} is less than
%% 50\%, so we are further fine-tuning the fault tolerance analysis and
%% QEC step insertion heuristics with the goal of reaching a 50\% success
%% rate for all synthesized circuits.

%% CQLA serializes QEC steps such that data must wait for ancilla
%% creation that begins once the data becomes idle.  Our first test slows
%% down execution of our circuit to the point where it matches CQLA.  At
%% this test point with equivalent latency, we find a significant
%% improvement in area.

%% However, this suboptimal latency is not necessarily the best point in
%% our design space.  So we next allow our CAD flow to find its own best
%% layout for the ???-bit factorization algorithm circuit.

%% Walk through:
%% \begin{itemize}
%% \item how our CAD flow can do this
%% \item what's automated, what's not
%% \item what's controlled online
%% \item what are tradeoffs in area, latency, error and online control
%%   complexity
%% \item focus on compute area reuse and minimal parallelism to maintain
%%   functionality of error correction and stuff
%% \end{itemize}

\section{Related Work}\label{sec:related}

We are not the first to build a computer-aided design flow for quantum
circuits.  Balensiefer, \etal~introduced one of the first complete
frameworks for a quantum CAD flow~\cite{eval_framework}.  Shende, \etal,
investigated logic optimization for quantum
circuits~\cite{shende2006sql}.  Since we focus on reducing circuitry
outside of the logical application circuit, our two approaches would be
complementary.  The concept of selective error correction was suggested
by Oskin, \etal~\cite{oskin2002par}, although they did not provide
guidance on how to perform these optimizations.

Draper pioneered the development of quantum adders~\cite{draper2000aqc,
draper2004ldq}, while Van Meter and Itoh investigated the impact of
communication and available concurrency on adder
performance~\cite{vanmeter2004fqm}.  Our adder partitioning experiments
work at a lower level, taking the logical adder structure as a given and
determining physical geometries that perform best.

Using a teleportation network for long range communication was proposed
in~\cite{oskin2002par} and further exploited in~\cite{metodi2005qla,
thaker2006qmh}.  Our work provides the first systematic attempt to size
router resources automatically based on circuit topology.

Although others have simulated error propagation in quantum
circuits~\cite{steane2003oan, viamontes2003gls, cross2007ccs}, we are
the first to provide detailed error simulation for a modular design.

Finally, Shor's factoring has been the focus of many studies.  Some
notable ones include~\cite{fowler2004sss, metodi2005qla,
vanmeter2004fqm}.  In particular, Metodi, \etal~\cite{metodi2005qla}
estimated design area for 1024-bit factoring at 0.9 $m^{2}$, using a
model of QLA.  Our work advances the state of the art by applying more
efficient resource allocation and significantly reducing the design
area.

\section{Conclusion}\label{sec:conclusion}

We examined quantum circuits by laying them out in a set of five
different ``datapath organizations''.  We optimized the area and latency
of Shor's factoring while simultaneously improving fault tolerance
through: (1) balancing the use of ancilla generators, (2) aggressive
optimization of error correction, and (3) tuning the core adder
circuits.  In the process, we introduced a metric, called ADCR, which is
the probabilistic equivalent of the classic Area-Delay product.  ADCR
provides a natural mechanism for optimizing layouts and subsequently
evaluating area-efficiency---allowing us to eliminate two of the five
datapath organizations immediately (QLA and CQLA), while leaving three
others for further evaluation (LQLA, CQLA+, and Qalypso).

Our error correction optimization reduces ADCR by more than an order of
magnitude, especially for less ancilla-efficient datapaths.
Further, through a detailed analysis of optimized circuits, we showed
that the area of a quantum circuit can have as little as 20\% devoted
exclusively to quantum error correction; this result belies conventional
wisdom.

Finally, since quantum addition is at the heart of Shor's factoring, we
explored quantum adder architectures, showing that QCLA adders beat
ripple-carry adders by a factor of 20 in ADCR, despite being larger and
more complex.  We concluded by mapping a complete Shor's factoring circuit
to show 7659 mm$^{2}$ for the smallest circuit and $6 \times 10^8$
seconds for the fastest circuit.

\section{Acknowledgments}

We would like to thank Mark Oskin and the reviewers for their helpful
suggestions on improving the paper.  Further, we would like to thank
Lucas Kreger-Stickles for his help in understanding the ancilla
generator used in LQLA.

{\small
%\bibliographystyle{unsrt}
%\vspace{0.12in}
\bibliographystyle{plain}
\bibliography{main}
%\bibliogrpahy{main2}
}
\end{document}